\documentclass[prb,twocolumn,superscriptaddress,showpacs,amsmath,amssymb,article]{revtex4}
\usepackage{amsfonts}
\usepackage{bbm}
\usepackage{graphicx}% Include figure files
\usepackage{dcolumn}% Align table columns on decimal point
\usepackage{bm}% bold math
\usepackage{subfigure}
\usepackage{color}

\usepackage[colorlinks=true, linkcolor=blue, urlcolor=blue, citecolor=blue]{hyperref}

\begin{document}
\title{Josephson effects in the spin-triplet superconductor/altermagnet/spin-triplet superconductor junctions: the detection of the intrinsic $\bf{d}$-vector}
\author{Ya-Ting Han}
\affiliation{School of Science, Qingdao University of Technology, Qingdao, Shandong 266520, China}

\author{Li-Juan Chen}
\affiliation{School of Science, Qingdao University of Technology, Qingdao, Shandong 266520, China}

\author{Wen-Ting Liu}
\affiliation{School of Science, Qingdao University of Technology, Qingdao, Shandong 266520, China}

\author{Qiang Cheng}
\email[]{chengqiang07@mails.ucas.ac.cn}
\affiliation{School of Science, Qingdao University of Technology, Qingdao, Shandong 266520, China}
\affiliation{International Center for Quantum Materials, School of Physics, Peking University, Beijing 100871, China}

\author{Qing-Feng Sun}
\email[]{sunqf@pku.edu.cn}
\affiliation{International Center for Quantum Materials, School of Physics, Peking University, Beijing 100871, China}
\affiliation{Hefei National Laboratory, Hefei 230088, China}

\begin{abstract}
We study the Josephson effects in the spin-triplet superconductor/altermagnet/spin-triplet superconductor junctions using the Green's function method. It is found that the current-phase difference relationships in the junctions strongly
depend on the direction of the $\bf{d}$-vectors in the spin-triplet superconductors and the orientation angle of the altermagnet. For the given orientation angle, the $0$-$\pi$ transition can be obtained when the $\bf{d}$-vector is rotated.
The variations of the critical current of the junctions with the direction of the $\bf{d}$-vector, the orientation angle and the strength of altermagnetism are systematically investigated. These Josephson effects can provide the distinguishable information about the direction of the $\bf{d}$-vector.
Compared to the existing research, the proposed altermagnetic Josephson junctions can effectively avoid the negative influence of the magnetic field on the $\bf{d}$-vector and can serve as a feasible scheme for the detection of the intrinsic $\bf{d}$-vector. The obtained $0$-$\pi$ transition in the junctions can also have potential applications in the design of quantum devices.
\end{abstract}

\maketitle

\section{\label{sec1}Introduction}

As the most promising candidate of the spin-triplet superconductor (STS), Sr$_{2}$RuO$_{4}$ has continuously attracted
tremendous interest in the past thirty years\cite{1} and has once
again become a hot topic in condensed matter physics
recently\cite{2,3,4,5,6,7}. Different from its singlet counterpart, the pairing wave function of STS can be described by the
so-called $\bf{d}$-vector which includes both orientation and
the orbital part\cite{8}. The determination of the $\bf{d}$-vector not
only can help to reveal the physical mechanism of the
spin-triplet pairing in STS but also can lay the foundation
for the realization of the topological quantum states\cite{9,10}.
However, the form of the orbital part and the direction
of the $\bf{d}$-vector in the candidate material of STS are still
full of controversy\cite{11}.

For Sr$_{2}$RuO$_{4}$, its $\bf{d}$-vector with the time-reversal breaking $p$-wave pairing has been proposed based on the
muon spin-relaxation measurements\cite{12}. The Knight-shift\cite{13}
measurements support the above conclusion and demonstrate that the chiral $p$-wave pairing is consistent with
the experimental results\cite{14}. In contrast, the subsequent
measurements of the spin-lattice relaxation rate and the
thermal conductivity measurements show that the orbital part of the $\bf{d}$-vector exhibits the highly anisotropic
structure\cite{15,16} and the vertical line nodes\cite{17}, which excludes
the possibility of the $p$-wave pairing. The recent nuclear magnetic resonance experiments and the observed
magnetic-field dependence of the Josephson critical current also contradict the chiral $p$-wave state in the superconducting Sr$_{2}$RuO$_{4}$\cite{18,19}.
In addition to the orbital
part discussed above, the direction of the $\bf{d}$-vector also remains to be determined. The polarized-neutron scattering studies and the phase-sensitive measurements in
Josephson junctions indicate that the direction of the
$\bf{d}$-vector is along the crystallographic $c$-axis\cite{20,21} while
the experiment in Ref.[\onlinecite{18}] rules out the $\bf{d}$-vector along
the $c$-axis. Furthermore, the reduction of the spin susceptibility observed in experiments is consistent with the
time-reversal symmetric helical $p$-wave pairing with the
$\bf{d}$-vector in the crystallographic $ab$-plane\cite{22}. This conclusion is supported by the transport properties of the
planar and corner Josephson junctions formed between
Sr$_{2}$RuO$_{4}$ and Nb\cite{23}.

In addition to the aforementioned experiments, numerous theoretical
schemes have been proposed to detect the $\bf{d}$-vector in
Sr$_{2}$RuO$_{4}$ \cite{24,25,26,27,28,29,30,31,32,33,34,35,36,37}.
For example, in the $c$-axis-oriented normal metal/ferromagnet/STS junctions,
the voltage dependence of the tunneling conductance can provide a signal for
distinguishing the chiral $p$-wave pairing and the helical
$p$-wave pairing when the $\bf{d}$-vector is parallel to the direction of magnetization\cite{29}.
In the $s$-wave superconductor/STS Josephson junctions, it is found that a $\cos\varphi$ term appears in the current-phase relation for the chiral
$p$-wave pairing while the term is absent for the helical
$p$-wave pairing\cite{31}. Ref.[\onlinecite{31}] also examines the distinct dependences of the critical current on the external magnetic flux for these two types of pairings.
Additionally, tunneling spectroscopy of three-dimensional normal metal/Sr$_{2}$RuO$_{4}$ junctions along the $a$- and $b$-axes can differentiate helical $p$-wave pairing from higher-harmonic pairing\cite{33}.
The inverse proximity effects at the
STS/ferromagnet interface are also investigated for the
chiral and helical pairings for different relative orientations of magnetization and the $\bf{d}$-vector, which can
provide the immediate detection signatures for the spin-triplet pairs\cite{36}.

So far, most of the experiments and theoretical researches on the determination of the $\bf{d}$-vector require the
assistance of the external field or the exchange field in
order to obtain the triplet pairing sensitive results\cite{12,13,14,15,17,18,19,20,21,22,23,24,25,28,29,31,32,33,35,36,37}. However, theoretical and experimental studies indicate that the direction of the $\bf{d}$-vector can be changed by a small field\cite{Annett,Murakawa}. Especially, the authors in Ref.[\onlinecite{14}] recently demonstrated the instantaneous destruction of superconductivity by the RF pulses
in their early experiments\cite{KIshida}. This makes the determination of the intrinsic $\bf{d}$-vector in STS even more elusive. The detection of the direction of the intrinsic $\bf{d}$-vector is still an open question in condensed matter physics. A non-destructive but $\bf{d}$-vector sensitive detection scheme is urgent.

Recently, a new class of magnetic order, altermagnet (AM), has attracted considerable attention due to its distinctive characteristics and potential applications\cite{38,39,41,42,43,49,50,51,52,53,Ghorashi,YXLi,Giil,add1,add2,add3,Herasymchuk,Yarmohammadi,Tjernshaugen,Sukhachov,add4,Hodt,Amundsen}. The Andreev reflection, the odd-frequency pairing, the Josephson effects and the superconducting diode effects are also considered in the altermagnetic junctions\cite{Papaj,46,47,57,45,56,Banerjee,55,Maeda}. For example, the charge and spin transports in the AM/$s$- or $d$-wave superconductor junctions are examined and their strong dependence on the orientation of AM is reported\cite{Papaj,46}. The $0$-$\pi$ transition and the nonsinusoidal current-phase relationship (CPR) are predicted in the altermagnetic Josephson junctions with the $s$-wave pairing. The decay length and the oscillation period of the Josephson coupling strongly depend on the crystallographic orientation of AM\cite{47,57}. The superconducting diodes based on AM are theoretically designed without the assistance of an external field\cite{56,Banerjee} and the exotic odd-frequency pair amplitudes with higher angular momentum are classified according to their symmetries\cite{55,Maeda}.

In this paper, we propose the STS/AM/STS Josephson junctions as a scheme to present the detection of the intrinsic $\bf{d}$-vector in STS. Compared to the existing studies, there are two significant advantages in our scheme. The first is that AM possesses the zero-net macroscopic magnetization\cite{41,42}, which effectively avoids the negative influence on the intrinsic $\bf{d}$-vector. The second is that AM has anisotropic spin-polarization\cite{41,42}, which not only ensures that the Josephson effects in the STS/AM/STS junctions are still sensitive to the direction of the $\bf{d}$-vector but also provides an extra degree of freedom for the investigation of the Josephson effects. For STS, we adopt the highly controversial chiral $p$-wave state as an example to demonstrate the effectiveness of our scheme. Through numerical calculations based on the Green's function method, the $\bf{d}$-vector direction-resolved CPRs can be achieved. The $0$-$\pi$ transition can happen when the $\bf{d}$-vector is rotated. For a fixed direction of the $\bf{d}$-vector, the CPRs depend on the crystallographic orientation of AM. The $0$-$\pi$ transition can also be realized. The invariances of the CPRs are analysed through the symmetry analysis. The behavior of the critical current is discussed when the direction of the $\bf{d}$-vector, the orientation angle or the altermagnetism strength is altered. A comprehensive physical explanation of the numerical results is provided, which includes perspectives such as the Andreev reflection process, the behavior of electrons and holes on anisotropic Fermi surfaces, and comparisons with the ferromagnetic Josephson junctions. These results are helpful in the determination of the intrinsic $\bf{d}$-vector and in the design of the quantum devices.

The structure of this paper is arranged as follows. Sec. \ref{sec2} introduces the model of the proposed STS/AM/STS junctions and presents the expression of the Josephson current with a detailed description of the discretized model. Sec. \ref{sec3} shows and discusses the numerical results, focusing on the CPRs and the variations in the critical current. Sec. \ref{sec4} provides a comprehensive physical explanation of the numerical results.
Sec. \ref{sec5} provides a summary of the paper. The Appendix gives a detailed derivation process of the discretization of each Hamiltonian, the corresponding Green's functions, the proof of the zero-net macroscopic magnetization of the AM and the numerical results for the ferromagnetic Josephson junctions.

\section{\label{sec2}Model and Formulation}

\begin{figure}[!htb]
	\centerline{\includegraphics[width=\columnwidth]{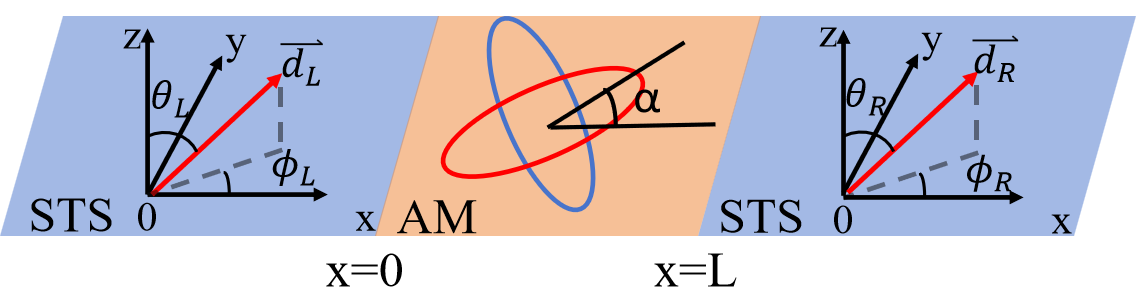}}
	\caption{ Schematic diagram of the STS/AM/STS Josephson junctions. The red arrow in the left (right) STS denotes the vector ${\bf{d}}_{L(R)}$. Its direction is characterized by the polar angle $\theta_{L(R)}$ and the azimuthal angle $\phi_{L(R)}$. The red and blue ellipses represent the Fermi surface in AM. The former is occupied by the spin-down electrons while the latter is occupied by the spin-up electrons. The orientation of AM is defined as the angle $\alpha$ between the major axis of the red ellipse and the interface normal. \label{fig1}}
\end{figure}

We consider the two-dimensional STS/AM/STS Josephson junctions in the $xy$-plane as shown in Fig.\ref{fig1}. The Josephson current flows along the $x$-axis and the two interfaces are located at $x=0$ and $x=L$. The length of AM is $L$, which is confined to the region of $0<x<L$. For AM, the $d$-wave altermagnetism is considered. Its orientation angle $\alpha$ is defined as the angle between the crystalline axis (the major axis of the red ellipse as shown in Fig.\ref{fig1}) and the normal direction of the interface. For $\alpha=0$, AM is of the pure $d_{x^2-y^2}$-wave altermagnetism while for $\alpha=\pi/4$, it is of the pure $d_{xy}$-wave altermagnetism. The $\bf{d}$-vector of the left (right) semi-infinite STS is denoted by ${\bf{d}}_{L(R)}=\Delta_{0}e^{i\varphi_{L(R)}}f({\bf{k}}){\bf{n}}_{L(R)}$ with the unit vector ${\bf{n}}_{L(R)} =\left(\sin\theta_{L(R)}\cos\phi_{L(R)},
\sin\theta_{L(R)}\sin\phi_{L(R)},\cos\theta_{L(R)}\right)$. Here, $\Delta_0$ is the gap magnitude, $\varphi_{L(R)}$ is the superconducting phase,
$\theta_{L(R)}$ and $\phi_{L(R)}$ are the polar angle and the azimuthal angle of the $\bf{d}$-vector, respectively. For definiteness, we choose the widely discussed state of Sr$_{2}$RuO$_{4}$, i.e., the chiral $p$-wave state with $f({\bf{k}})=k_x+ik_y$, as an example to provide the Josephson information for the direction of the ${\bf{d}}$-vector. Since both are Sr$_{2}$RuO$_{4}$, we take ${\bf{d}}_L\parallel{\bf{d}}_{R}$. In other words, we have $\theta_L=\theta_R=\theta$ and $\phi_{L}=\phi_R=\phi$.

The Hamiltonian of the junctions can be written as\cite{45,56}
\begin{eqnarray}
{H}=H_{L}+H_{AM}+H_{R},\label{1}
\end{eqnarray}
where $H_{L}$,$H_{AM}$ and $H_{R}$ represent the Hamiltonians of the left STS, the AM and the right STS, respectively.

The Hamiltonian $H_{L(R)}$ for the left (right) STS is given by
\begin{eqnarray}
H_{L(R)}=\sum_k\psi_{L(R)\textbf{k}}^{\dagger}\check{H}_{L(R)}(\textbf{k})\psi_{L(R)\textbf{k}},\label{2}
\end{eqnarray}
with $\psi_{L(R){\textbf{k}}}=(c_{L(R),\textbf{k}\uparrow},c_{L(R),\textbf{k}\downarrow},c_{L(R),\textbf{-k}\uparrow}^{\dagger},c_{L(R),\textbf{-k}\downarrow}^{\dagger})^T$ and the $4\times4$ Bogoliubov-de Gennes (BdG) Hamiltonian is given by
\begin{eqnarray}
	\check{H}_{L(R)}(\textbf{k})=\left(\begin{array}{cc}
		h_{L(R)}(\textbf{k})&\Delta_{L(R)}(\textbf{k})\\
		-\Delta_{L(R)}^*(-\textbf{k})&-h_{L(R)}^*(-\textbf{k})
	\end{array}\right).\label{3}
\end{eqnarray}
Here, $h_{L(R)}(\textbf{k})=t_{0}\textbf{k}^{2}-\mu_{L(R)}$ with the chemical potential $\mu_{L(R)}$ and $\Delta(\textbf{k})=({\bf{\sigma}}\cdot\textbf{d}_{L(R)})\mathit{i}\sigma_{y}$ with the Pauli matrix ${\bf{\sigma}}$=($\sigma_{x}$,$\sigma_{y}$,$\sigma_{z}$). For AM, we adopt the following  Hamiltonian\cite{45,56}
\begin{eqnarray}
	H_{AM}=\sum_k\psi_{\bf{k}}^{\dagger}\check{H}_{AM}(\bf{k})\psi_{\bf{k}},\label{4}
\end{eqnarray}
with $\psi_{\bf{k}}=(c_{\bf{k}\uparrow},c_{\bf{k}\downarrow},c_{\bf{-k}\uparrow}^{\dagger},c_{\bf{-k}\downarrow}^{\dagger})^T$ and the BdG Hamiltonian is given by
\begin{eqnarray}
	\check{H}_{AM}(\bf{k})=\left(\begin{array}{cc}
		h_{AM}(\bf{k})&0\\
		0&-h_{AM}^*(-\bf{k})
	\end{array}\right).\label{5}
\end{eqnarray}
Here $ h_{AM}(\textbf{k}) = [t_{0}(k_{x}^{2} + k_{y}^{2}) - \mu]\sigma_0 + t_{J}[(k_{x}^{2} - k_{y}^{2})\cos 2\alpha + 2k_{x}k_{y}\sin 2\alpha]\sigma_{z}$. The parameter $t_{J}$ characterizes the strength of the altermagnetism in AM and $\mu$ is the chemical potential in AM. The N$\acute{\textbf{e}}$el vector in AM is along the $z$-axis.

We discretize the above continuum Hamiltonians on a two-dimensional square lattice as shown in Fig.\ref{2} to calculate the Josephson current. The lattice constant is taken as $a$. The length and the width of the AM region can be expressed as $L=(N_{x}-1)a$ and $W=(N_{y}-1)a$, respectively. The discrete Hamiltonian for left (right) STS is given by
\begin{eqnarray}
	\begin{split}
 H_{L(R)}&=\sum_i[\psi_{i}^{L(R)\dagger}\check{H}_{0}^{L(R)}\psi_{i}^{L(R)}+\psi_{i}^{L(R)\dagger}\check{H}_{x}^{L(R)}\psi_{i+\delta{x}}^{L(R)}\\
	&{}+\psi_{i}^{L(R)\dagger}\check{H}_{y}^{L(R)}\psi_{i+\delta{y}}^{L(R)}+H.c.],
	\end{split}
	\label{6}
\end{eqnarray}
with $\psi_{i}^{L(R)}=(\psi_{L(R),i\uparrow},\psi_{L(R),i\downarrow},\psi_{L(R),i\uparrow}^{\dagger},\psi_{L(R),i\downarrow}^{\dagger})$. In the two-dimensional lattice, the subscript $\textbf{i}=(i_{x},i_{y})$ denotes the coordinate position of the \textbf{i}th lattice point and $i_y$ is limited in the region $1\leq i_{y}\leq N_{y}$. $i+\delta_x$ denotes the nearest-neighbor sites along the $x$ direction and $i+\delta_y$ denotes the nearest-neighbor sites along the $y$ direction.
For the STSs on the left and right sides, the range of $i_x$ is set to $i_{x}\leq 0$ and $i_{x}\geq N_{x}+1$, respectively.

\begin{figure}[!htb]
	\centerline{\includegraphics[width=\columnwidth]{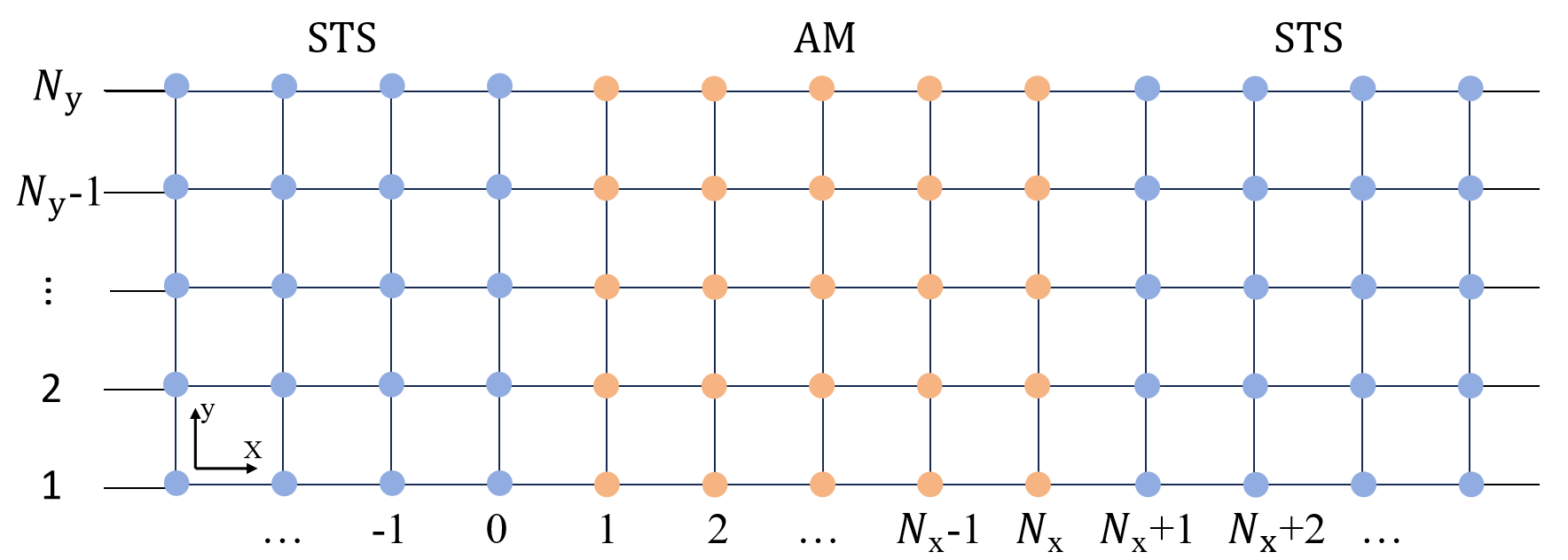}}
	\caption{Schematic illustration of the two-dimensional square lattice model for the discretization of the STS/AM/STS Josephson junctions.\label{fig2}}
\end{figure}
The discrete Hamiltonian for the middle AM can be given by
\begin{eqnarray}
	\begin{split}
		H_{AM}&=\sum_i\left[\psi_{i}^{\dagger}\check{H}_{0}^{AM}\psi_{i}
+\psi_{i}^{\dagger}\check{H}_{x}^{AM}\psi_{i+\delta{x}}
+\psi_{i}^{\dagger}\check{H}_{y}^{AM}\psi_{i+\delta{y}} \right.\\
&{}\left.+\psi_{i}^{\dagger}\check{H}_{xy}^{AM}\psi_{i+\delta_x+\delta_y}
+\psi_{i}^{\dagger}\check{H}_{x\bar{y}}^{AM}\psi_{i+\delta_x-\delta_y}
			+H.c.\right].
	\end{split}
	\label{7}
\end{eqnarray}
with $\psi_{i}=(\psi_{i\uparrow},\psi_{i\downarrow},\psi_{i\uparrow}^{\dagger},\psi_{i\downarrow}^{\dagger})$. In this region, $i+\delta_x+\delta_y$ and $i+\delta_x-\delta_y$
represent the next-nearest-neighbor sites of the \textbf{i}th site.
For AM, the range of $i_x$ is constrained to
$1 \leq i_x\leq N_x$ and that of $i_y$ to $1 \leq i_y \leq N_y$. The explicit form of the matrices $\check{H}_0^{L(R)}$, $\check{H}_x^{L(R)}$, $\check{H}_y^{L(R)}$, $\check{H}_0^{AM}$, $\check{H}_x^{AM}$, $\check{H}_y^{AM}$,
$\check{H}_{xy}^{AM}$ and $\check{H}_{x\bar{y}}^{AM}$ will be presented in Appendix \ref{A}.

The tunneling Hamiltonian describing hopping between the left STS and AM and between the right STS and AM can be written as
\begin{eqnarray}
	\begin{split}
H_{T}&=\sum_{i_y}[\psi_{(0,i_y)}^{L\dagger}\check{T}\psi_{(1,i_y)}+\psi_{(N_x+1,i_y)}^{R\dagger}\check{T}\psi_{(N_x,i_y)}],
	\end{split}
	\label{8}
\end{eqnarray}
with the hopping matrix $\check{T}=(t,t,-t^*,-t^*)$, where \textit{t} is a real number.  $(0,i_y)$, $(1,i_y)$, $(N_x+1,i_y)$ and $(N_x,i_y)$ denote the coordinates of the rightmost column of lattice points in the left STS, the leftmost column of lattice points in the AM, the leftmost column of lattice points in the right STS, and the rightmost column of lattice points in the AM, respectively.

We can define the particle number operator for the left STS as
\begin{eqnarray}
N=\sum_{\substack {i_x\le 0 \\ 1\leq i_y\leq N_y}}\sum_{\sigma=\uparrow,\downarrow}\psi_{(i_x,i_y)\sigma}^{+}\psi_{(i_x,i_y)\sigma}.
	\label{9}
\end{eqnarray}

Using the Green's function method\cite{37,Sunjpcm,YHLi}, the Josephson current can be expressed as
\begin{equation}
\begin{aligned}
I&=e\langle \frac{dN}{dt}\rangle
=-\frac{e}{2\pi}\int dE \text{Tr}[\Gamma_z\check{T}G^{<}(E)+H.c.],
\label{10}
\end{aligned}
\end{equation}
with $\Gamma_{z}=\sigma_z\otimes 1_{2\times2}$. The derivation of the lesser Green function $G^{<}(E)$ will be discussed in Appendix \ref{B}.
\section{\label{sec3}Numerical results and discussions}
Before discussing the numerical results, we emphasize that AMs possess zero-net macroscopic magnetization. This is an important advantage over ferromagnets in terms of constructing superconducting heterostructures. The zero-net macroscopic magnetization can effectively avoid the competition between magnetism and superconductivity and the negative influence of possible stray field on the ${\bf{d}}$-vector. The absence of the net macroscopic magnetization can be proved through symmetry analysis, which is provided in Appendix \ref{C}.

In our calculations, we take $t_0=1$ as the energy unit and $\mu_{L}=\mu_{R}=\mu=2$. For the lattice parameters, we have $N_x=N_y=10$ and $a=1$. The magnitude of the hopping between the left (right) STS and AM is taken as $t = 1$. Due to the spin rotation symmetry about the $z$-axis obeyed by our junctions, the Josephson current is irrespective of the azimuthal angle $\phi$ and we will choose $\phi=0.5\pi$ for definiteness. The unit of the Josephson current is taken as $e\Delta_{0}/2\pi$ with $\Delta_{0}=0.01$. Next, we discuss the numerical results of the Josephson effect in two subsections, i.e., the CPRs and the critical current, to demonstrate the detection of the direction of the $\bf{d}$-vector.

\subsection{CPRs}\label{A}
\begin{figure*}[!htb]
	\centerline{\includegraphics[width=2\columnwidth]{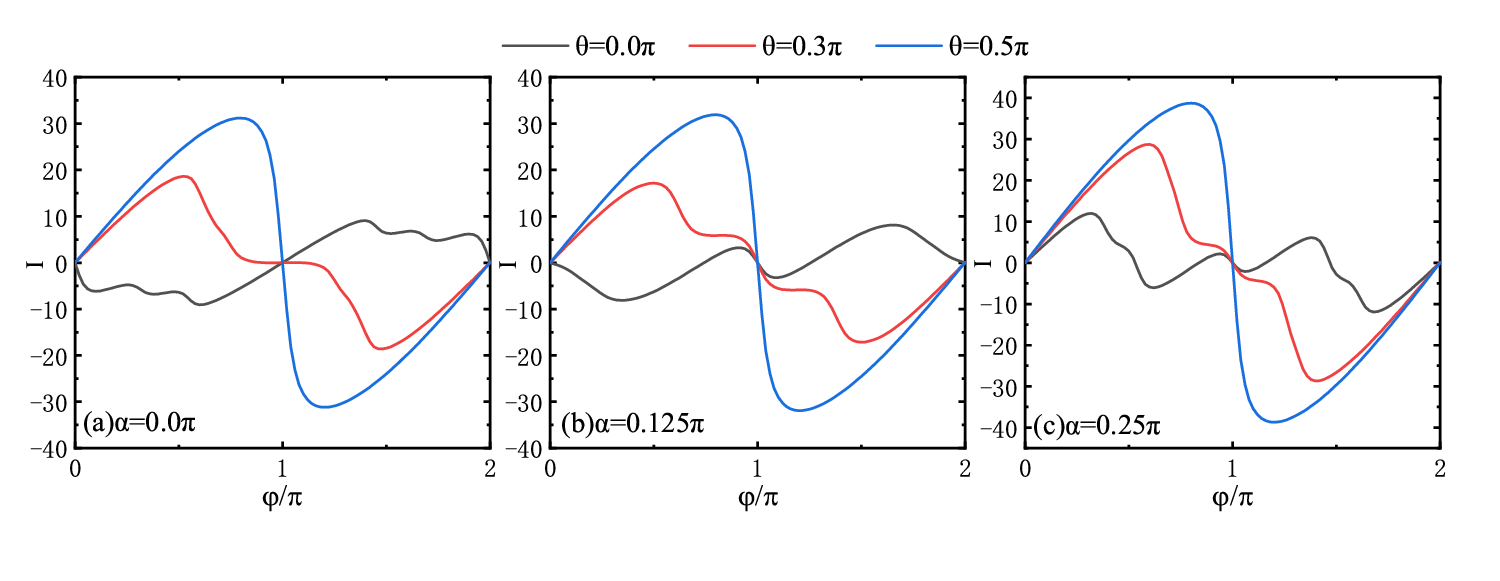}}
	\caption{ The CPRS for different polar angles with (a) $\alpha$=0.0$\pi$, (b) $\alpha$=0.125$\pi$ and (c) $\alpha$=0.25$\pi$. The strength of altermagnetism is taken as $t_J=0.5$.
The black, red, and blue solid lines correspond to $\theta  = 0$, $0.3\pi$, and $0.5\pi$, respectively.
\label{fig3}}
\end{figure*}

We present the results of CPRs for different orientation angles with fixed altermagnetism strength $t_J=0.5$ in Fig.\ref{fig3}. For a given orientation angle, the CPRs are strongly dependent on the direction of the $\bf{d}$-vector. In previous studies, the controversy over the direction of the ${\bf{d}}$-vector centers on whether it is in-plane or out-of-plane\cite{18,20,21}. Our results can provide clear information in clarifying this point. For $\alpha=0$ and $\alpha=0.125\pi$, the altermagnetic Josephson junction is in the $\pi$ state for the out-of-plane ${\bf{d}}$-vector with $\theta=0$ while it is in the $0$ state for the in-plane ${\bf{d}}$-vector with $\theta=0.5\pi$. The $0$-$\pi$ transition happens when the $\bf{d}$-vector is rotated from the in-plane direction to the out-of-plane direction. For a larger value of the orientation angle with $\alpha=0.25\pi$, the $0$-$\pi$ transition disappears. However, the strong dependence of CPRs on the direction of the $\bf{d}$-vector still survives in this situation. In other words, the CPRs in our junctions can provide the helpful information for detecting the direction of the $\bf{d}$-vector.

It is interesting to compare the altermagnetic junctions here with the ferromagnetic junctions. The interplay of ferromagnetism and the spin-triplet superconductivity in the latter junctions and its potential applications have been studied theoretically in various structures\cite{Kastening,Brydon1,Brydon2,Brydon4}. The $0$-$\pi$ transition is also predicted in the ferromagnetic Josephson structure with the spin-triplet pairing\cite{Brydon2}. However, the realization of the $0$-$\pi$ transition in our junctions possesses two important differences compared with the existing research in Ref.[\onlinecite{Brydon2}]. First, the zero-net macroscopic magnetization in AM not only avoids influencing the intrinsic $\bf{d}$-vector in STS but also effectively prevents the magnetic cross-talk in the design of quantum devices\cite{Jungwirth}. Second, distinct from the isotropic spin-polarization in ferromagnet, the spin-polarization in AM is anisotropic as shown in Fig.\ref{fig1} and it provides a new degree of freedom, i.e., the orientation angle $\alpha$, for the achievement of the $0$-$\pi$ transition as shown in Fig.\ref{fig3}. In a word, the $0$-$\pi$ transition realized in our junctions has more advantages in the realistic applications than that in the ferromagnetic junctions.

For a given orientation angle of AM, the CPRs in our junctions satisfy the following symmetry relation
\begin{eqnarray}
I(\theta,\phi,\alpha,\varphi)=I(\pi-\theta,\phi,\alpha,\varphi).\label{sr1}
\end{eqnarray}
That's why we don't show the CPRs for $0.5\pi<\theta\le\pi$ in Fig.\ref{fig3}. The symmetry relation in Eq.(\ref{sr1}) can be derived from the symmetry analysis. We introduce the gauge transformation $U_{1}(\eta)$ with $\eta=\frac{\pi}{2}$. Under the operation $U_{1}(\frac{\pi}{2})$, the Hamiltonian of AM in Eq.(\ref{4}) remains unchanged while the Hamiltonian of the left (right) STS becomes\cite{37}
\begin{eqnarray}
\begin{split}
U_{1}(\frac{\pi}{2}) H_{L(R)}(\theta,\phi,\varphi_{L(R)})U_{1}(\frac{\pi}{2})^{-1}
\\=H_{L(R)}(\pi-\theta,\pi+\phi,\varphi_{L(R)}).
\end{split}
\end{eqnarray}
Since the operation $U_{1}(\frac{\pi}{2})$ is a unitary transformation, the Josephson current is invariant and one will have $I(\theta,\phi,\varphi)=I(\pi-\theta,\pi+\phi,\varphi)$. Because the Josephson current is irrespective of the azimuthal angle $\phi$, the symmetry relation in Eq.(\ref{sr1}) is obtained.

For the given polar and azimuthal angles of the $\bf{d}$-vector, we also have the following symmetry relations
\begin{eqnarray}
I(\alpha,\varphi)=I(\frac{\pi}{2}+\alpha,\varphi),\label{sr2}
\end{eqnarray}
and
\begin{eqnarray}
I(\alpha,\varphi)=I(\pi-\alpha,\varphi).\label{sr3}
\end{eqnarray}
For the relation in Eq.(\ref{sr2}), we introduce the spin-rotation $R_{y}(\pi)$ about the $y$-axis with the angle $\pi$. Under the operation, the Hamiltonians transform as follows\cite{37}
\begin{eqnarray}
R_{y}(\pi)H_{AM}(\alpha)R_{y}(\pi)^{-1}=H_{AM}(\frac{\pi}{2}+\alpha),
\end{eqnarray}
and
\begin{eqnarray}
\begin{split}
R_{y}(\pi)H_{L(R)}(\theta,\phi,\varphi_{L(R)})R_{y}(\pi)^{-1}\\
=H_{L(R)}(\pi-\theta,\pi-\phi,\varphi_{L(R)}).
\end{split}
\end{eqnarray}
Due to the unitary property of $R_{y}(\pi)$, the Josephson current is invariant under the transformation, which implies $I(\theta,\phi,\alpha,\varphi)=I(\pi-\theta,\pi-\phi,\frac{\pi}{2}+\alpha,\varphi)$.
Combing the relation in Eq. (\ref{sr1}) and the fact that the Josephson current is independent of the azimuthal angle $\phi$, we derive the relation in Eq.(\ref{sr2}).

For the relation in Eq.(\ref{sr3}), we introduce the joint operation $X=TM_{yz}$, where $M_{yz}$ is the mirror reflection about the $yz$-plane and $T$ is the time-reversal operation. Under the joint operation, the Hamiltonians transform as follows\cite{37}
\begin{eqnarray}
XH_{AM}(\alpha)X^{-1}=H_{AM}(\pi-\alpha),
\end{eqnarray}
and
\begin{eqnarray}
%\begin{split}
XH_{L(R)}(\theta,\phi,\varphi_{L(R)})X^{-1}
=H_{R(L)}(\theta,\pi-\phi,-\varphi_{L(R)}).
%\end{split}
\end{eqnarray}
The mirror operation $M_{yz}$ exchanges the left STS and the right STS and the Josephson current is reversed. However, the subsequent time-reversal operation will change the reversed Josephson current back to the original direction. Then, one has $I(\theta,\phi,\alpha,\varphi)=I(\theta,\pi-\phi,\pi-\alpha,\varphi)$ which will lead to the symmetry relation in Eq.(\ref{sr3}).

Because of the establishments of Eqs.(\ref{sr1}),(\ref{sr2}) and (\ref{sr3}), we only consider CPRs for the orientation angle $0\le\alpha\le0.25\pi$ in Fig.\ref{3}. The CPRs for $0.25\pi<\alpha<2\pi$ can be obtained from the symmetry relations in Eqs.(\ref{sr1}),(\ref{sr2}) and (\ref{sr3}). It is worth emphasizing that the dependence of CPRs on the orientation angle $\alpha$ is a significant transport feature peculiar to the altermagnetic junctions. For the STS/ferromagnet/STS junctions, this feature is not expected due to the isotropic spin-polarization in ferromagnets. The involvement of AM in Josephson junctions provides a new degree of freedom for the detection of the $\bf{d}$-vector in STS.
\subsection{Critical current}\label{B}
\begin{figure}[!htb]
\centerline{\includegraphics[width=\columnwidth]{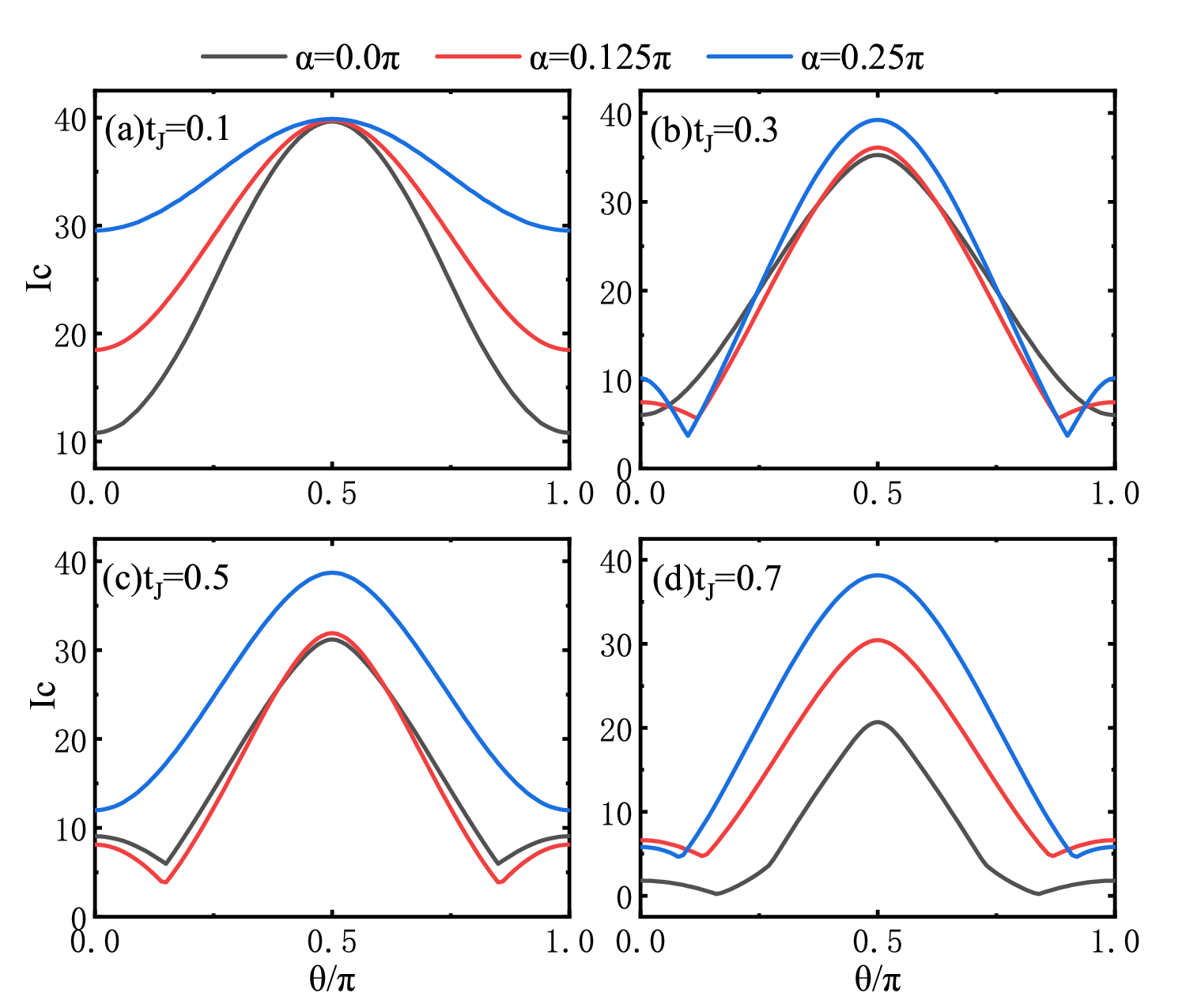}}
\caption{The critical current as a function of the polar angle of the $\bf{d}$-vector for different orientation angles with (a) $t_J=0.1$, (b) $t_J=0.3$, (c) $t_J=0.5$ and (d) $t_J=0.7$. The black, red, and blue solid lines correspond to $\theta  = 0$, $0.3\pi$, and $0.5\pi$, respectively. \label{fig4}}
\end{figure}

The critical current $I_{c}$ is defined as $I_{c}=\text{max}(\vert I\vert)$. In Fig.\ref{fig4}, we present the variations of $I_c$ as the polar angle of the $\bf{d}$-vector for different altermagnetism strengths and orientation angles.
First, the critical current is symmetric about the polar angle $\theta=0.5\pi$ for various situations, which is consistent with the symmetry relation in Eq.(\ref{sr1}). Second, the critical current is strongly dependent on the polar angle, i.e., the direction of the $\bf{d}$-vector, for each situation with the given values of $t_J$ and $\alpha$.
In other words, the critical current can provide effective information of the direction about the $\bf{d}$-vector. For the small altermagnetism strength with $t_{J}=0.1$ as shown in Fig.\ref{fig4}(a), the curves of $I_{c}$ are smooth for different orientation angles, which means that there is no $0$-$\pi$ transition in this situation. Furthermore, as AM is changed from the $d_{x^2-y^2}$-wave AM with $\alpha=0$ to the $d_{xy}$-wave AM with $\alpha=0.25\pi$, the critical current is significantly increased except for a small angle range near $\theta=0.5\pi$. For the moderate altermagnetism strength such as $t_{J}=0.3$ in Fig.\ref{fig4}(b) or $t_{J}=0.5$ in Fig.\ref{fig4}(c), dips appear in the critical current curves, which implies that the $0$-$\pi$ transitions can happen in this situation. The $0$-$\pi$ transitions happen when $\alpha$ is not near $0$ for $t_{J}=0.3$ while they happen when $\alpha$ is not near $0.25\pi$ for $t_J=0.5$. However, for the larger altermagnetism strength (e.g., $t_{J}=0.7$), the $0$-$\pi$ transition can happen for all orientation angles as shown in Fig.\ref{fig4}(d). For example, for $\alpha=0.25\pi$, the minimum point of the blue line is at $\theta_0\approx0.92\pi$. For $\theta<\theta_0$, the junctions are in the $0$ state and for $\theta>\theta_0$, the junctions are in the $\pi$ state. The $0$-$\pi$ transition occurs at $\theta=\theta_0$. This is why the critical current is enhanced at $\theta=\pi$ at $\alpha=0.25\pi$. Specifically, as AM is changed from the $d_{x^2-y^2}$ AM with $\alpha=0$ to the $d_{xy}$-wave AM with $\alpha=0.25\pi$, the critical current demonstrates a marked enhancement over a broad angular range around $\theta=0.5\pi$. For a general ${\bf{d}}$-vector that is neither in-plane nor out-of-plane, our results can also provide helpful information about its orientation as shown in Fig.\ref{fig4}. For example, the $0$-$\pi$ transition occurs at $\theta\approx0.15\pi$ for $\alpha=0.125\pi$ in Fig.\ref{fig4}(c), which indicates that the altermagnetic Josephson junction is in the $\pi$ state for the polar angle $\theta<0.15\pi$ while it is in the $0$ state for $\theta>0.15\pi$.

\begin{figure}[!htb]
	\centerline{\includegraphics[width=\columnwidth]{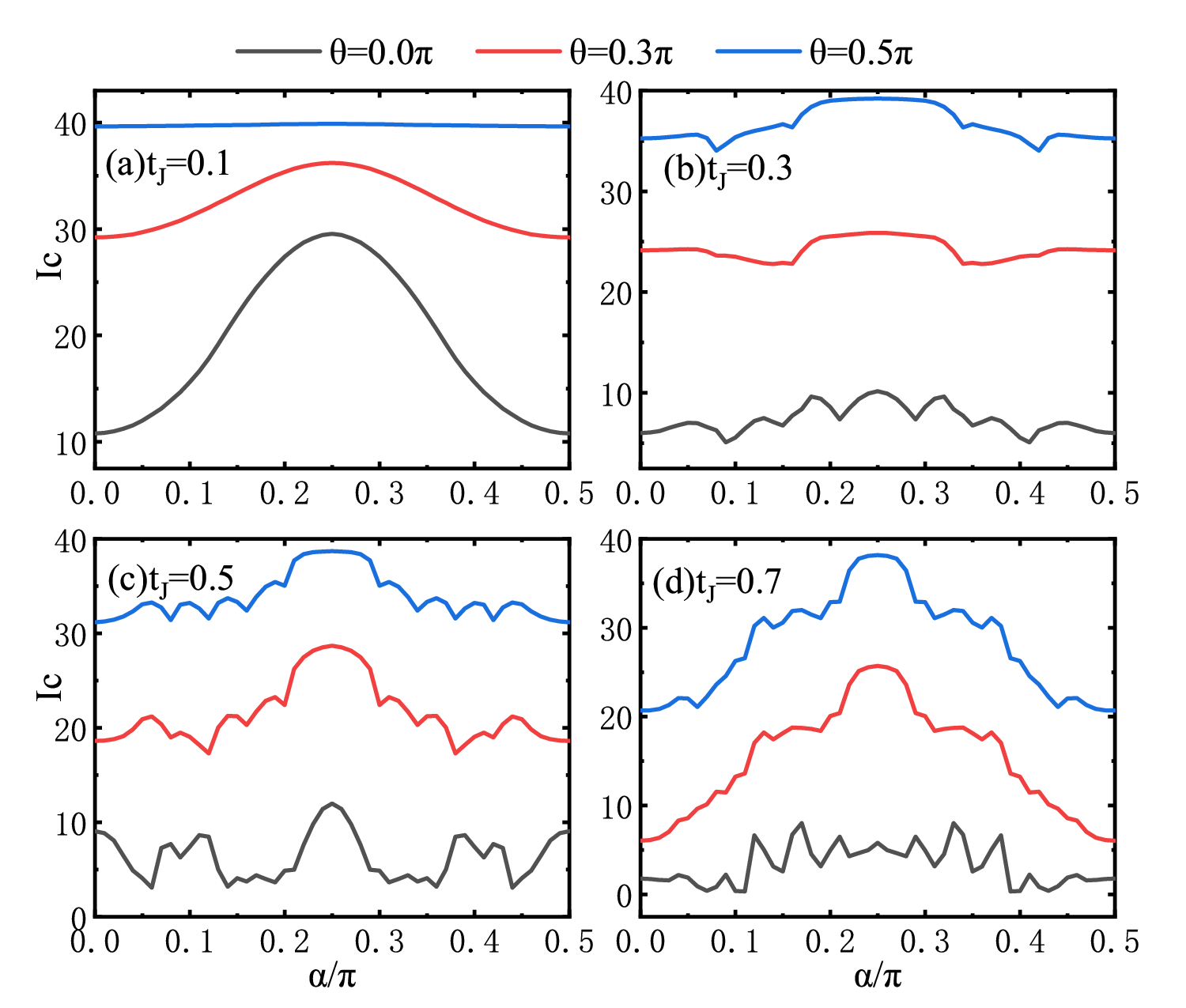}}
	\caption{The critical current as a function of the orientation angle $\alpha$  of AM for different polar angles with (a) $t_J=0.1$, (b) $t_J=0.3$  (c) $t_J=0.5$ and (d) $t_J=0.7$. The black, red, and blue solid lines correspond to $\theta  = 0$, $0.3\pi$, and $0.5\pi$, respectively. }\label{fig5}
\end{figure}

In Fig.\ref{fig5}, we present the variations of the critical current as the orientation angle $\alpha$ of AM for different altermagnetism strengths and polar angles of the $\bf{d}$-vector.
The critical current is symmetric about $\alpha=0.25\pi$ as shown in Fig.\ref{fig5} due to the relation $I(\alpha)=I(\frac{\pi}{2}-\alpha)$, which can be immediately deduced from Eqs.(\ref{sr2}) and (\ref{sr3}). For the small altermagnetism strength with $t_{J}=0.1$, the curves of the critical current are smooth as shown in Fig.\ref{fig5}(a) because of the absence of the $0$-$\pi$ transition in this situation. This feature of the critical current is consistent with that in Fig.\ref{fig4}(a). Furthermore, as the polar angle of the $\bf{d}$-vector is decreased from $\theta=0.5\pi$ to $\theta=0$, the dependence of the critical current on the orientation angle becomes stronger.
For $\theta=0.5\pi$, i.e., $\bf{d}$-vector along the in-plane direction, the critical current remains almost constant as $\alpha$ is varied while for $\theta=0.5\pi$, i.e., $\bf{d}$-vector along the out-of-plane direction, the critical current exhibits a strong dependence on $\alpha$. This character is also consistent with the result in Fig.\ref{fig4}(a), which provides the distinguishable signal for the direction of the $\bf{d}$-vector. Therefore, our junctions can be used to effectively detect the direction of the intrinsic $\bf{d}$-vector. For larger altermagnetism strengthes, the critical current exhibits the oscillation behavior with the change of the orientation angle as shown in Figs.\ref{fig5}(b)-(d) due to the occurrence of the $0$-$\pi$ transition,which is consistent with the results in Figs.\ref{fig4}(b)-(d). When $t_{J}=0.7$, the oscillation of the critical current becomes more intense as the polar angle of the $\bf{d}$-vector decreases.
\begin{figure*}[!htb]
	\centerline{\includegraphics[width=2\columnwidth]{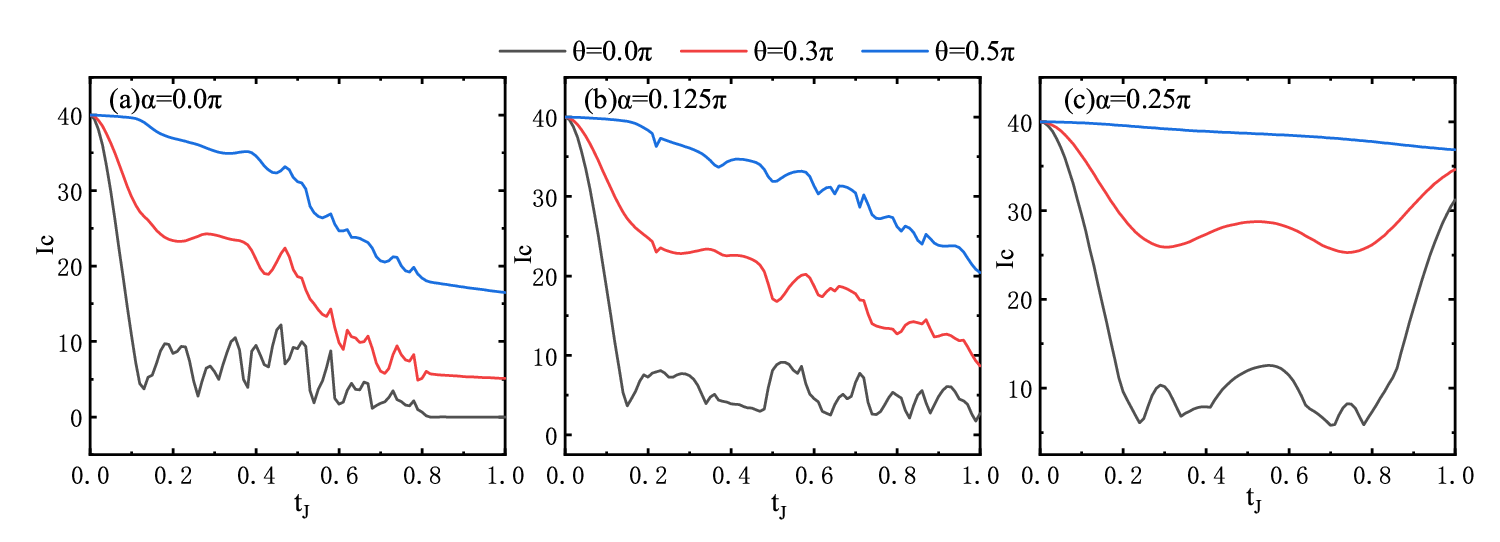}}
	\caption{The critical current as a function of the altermagnetism strength for different polar angles with (a) $\alpha=0.0\pi$, (b) $\alpha=0.3\pi$ and (c) $\alpha=0.5\pi$. The black, red, and blue solid lines correspond to $\theta  = 0$, $0.3\pi$, and $0.5\pi$, respectively. }\label{fig6}
\end{figure*}

In Fig.\ref{fig6}, we present the variations of the critical current as a function of the altermagnetism strength for different orientation angles and the polar angles of the $\bf{d}$-vector. For the orientation angle $\alpha=0$ or $\alpha=0.125\pi$, the critical current is obviously suppressed as the altermagnetism strength $t_{J}$ is enhanced as shown in Figs.\ref{fig6}(a) and (b). The obvious oscillation can also be observed due to the formation of the $0$-$\pi$ transition. However, for $\alpha=0.25\pi$, the curves of the critical current undergo a process from decreasing to increasing for $\theta=0$ and $\theta=0.3\pi$ and the curve for $\theta=0.5\pi$ is almost flat. The irregular oscillations in Figs.\ref{fig6}(a) and (b) become weaker for $\alpha=0.25\pi$ in Fig.\ref{fig6}(c). In addition, for the polar angles considered in Figs.\ref{fig5} and \ref{fig6}, the critical current is always increased when the polar angle is raised from $0$ to $0.5\pi$. These characters of the critical current can offer helpful information for the determination of the direction of the $\bf{d}$-vector.

\section{\label{sec4}Physical Explanation}
\begin{figure}[!htb]
	\centerline{\includegraphics[width=1\columnwidth]{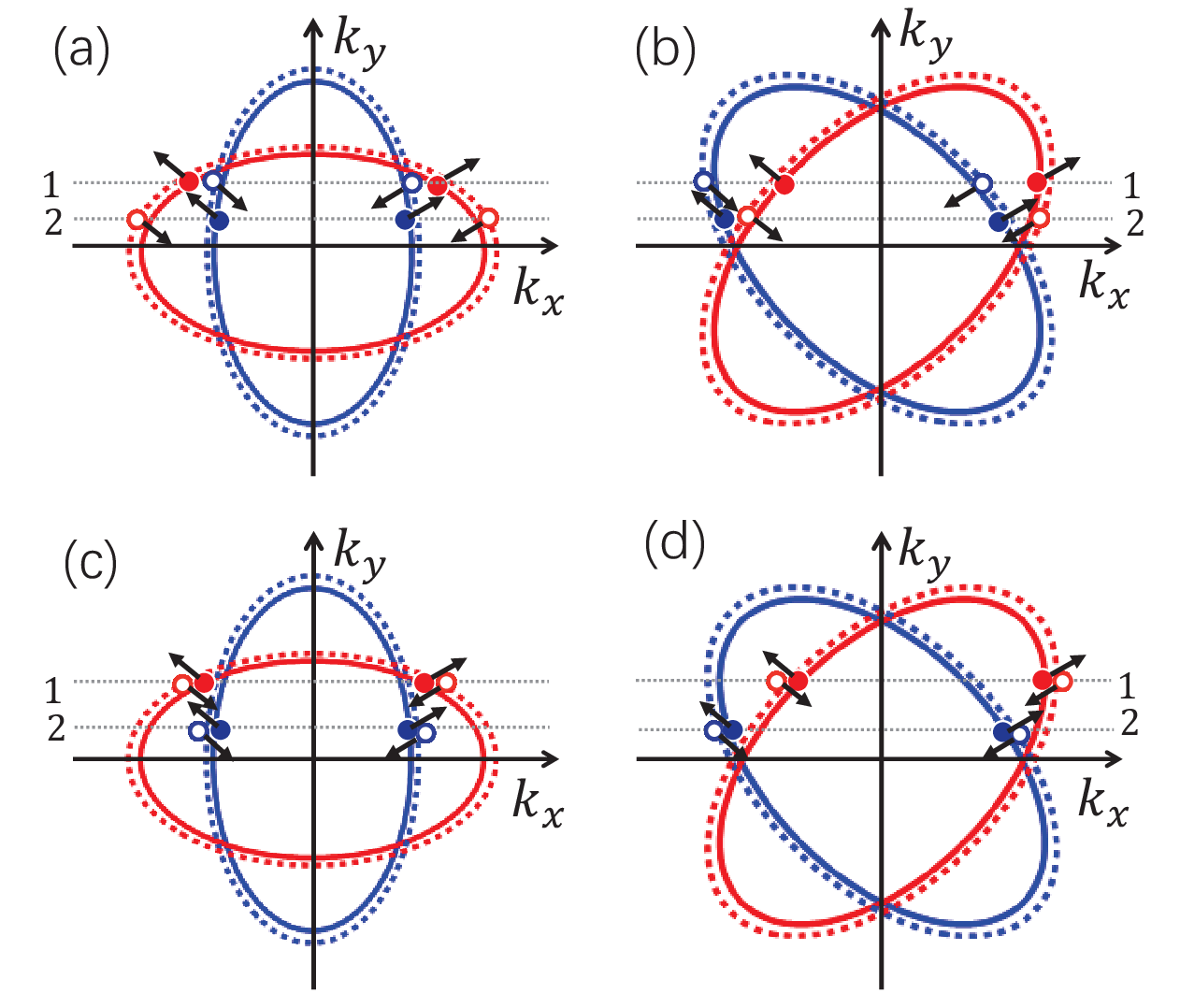}}
	\caption{The scattering processes on the anisotropic Fermi surfaces of AM for (a) the $d_{x^2-y^2}$-wave altermagnetism and the out-of-plane ${\bf{d}}$-vector, (b) the $d_{xy}$-wave altermagnetism and the out-of-plane ${\bf{d}}$-vector, (c) the $d_{x^2-y^2}$-wave altermagnetism and the in-plane ${\bf{d}}$-vector and (d) the $d_{xy}$-wave altermagnetism and the in-plane ${\bf{d}}$-vector. The red (blue) solid circles denote the spin-down (up) electrons and the red (blue) hollow circles denote the spin-down (up) holes. The black arrows denote the group velocities of particles. The electron and hole Fermi surfaces, which originally coincide, are drawn separately for the sake of distinction.}\label{fig7}
\end{figure}
For $\theta=0.5\pi$, i.e., the ${\bf{d}}$-vector being along the in-plane direction, the Cooper pairs in STSs are in the state with $S=1$ and $S_z=\pm 1$. Here, $S$ is the total spin of a Cooper pair and $S_z$ is its projection on the $z$-axis. Since the N$\acute{\text{e}}$el vector in the AM is along the $z$-axis, the spins of electrons in the AM are parallel to the spin projection $S_z$ of the equal-spin Cooper pairs in STSs. The tunneling of a single Cooper pair dominates the Josephson effect and the CPR for $\theta=0.5\pi$ is of the $\sin{\phi}$ form as shown in Fig.\ref{fig3}. For $\theta=0$, i.e., the ${\bf{d}}$-vector being along the out-of-plane direction, the Cooper pairs in STSs are in the state with $S=1$ and $S_{z}=0$. The spins of electrons in the AM are perpendicular to the spin projection $S_z$ of Cooper pairs in STSs. The coherent tunneling of two Cooper pairs is enhanced and the CPR for $\theta=0$ has a pronounced $\sin{2\phi}$ component, as shown in Fig.\ref{fig3}. For $0<\theta<0.5\pi$, the Cooper pairs in STSs are in the superposition state of the $S_z=0$ state and the $S_z=\pm 1$ state. The $\sin{2\phi}$ component in the CPR is reduced compared to the case of $\theta=0$ but remains visible, as shown in Fig.\ref{fig3}. These features of CPRs here are the same as those of ferromagnetic Josephson junctions when the magnetization in ferromagnet is along the $z$-axis (see Fig.D1(a) in Appendix D).

The Andreev reflection at the interfaces can convert electrons into holes or holes into electrons. The motion of electrons and holes can form a closed path. The Andreev bound states are formed when the phase accumulated on the closed path equals a multiple of $2\pi$. These bound states contribute to the Josephson current. For $\theta=0$, the Andreev reflection at the two interfaces converts the incoming spin-down (up) electron from the AM into the spin-up (down) hole entering the AM, and vice versa. The Andreev reflection processes are schematically shown in Figs.\ref{fig7}(a) and (b) by the circles on the gray dashed lines labeled ``1(2)''. The electrons and the Andreev reflected holes have the opposite spin. They occupy the identical ellipses with mutually perpendicular major axes, as shown in Figs.\ref{fig7}(a) and (b). As $t_J$ increases from $0$, the wave vectors of the electrons and the Andreev reflected holes become unequal, which leads to the rapid decay and oscillation of the critical current as shown by the black lines in Fig.\ref{fig6}. For $\alpha=0$, the spin-up electrons with $\vert k_y\vert>a_0$ cannot participate in the Andreev reflection due to the absence of the spin-down holes, as shown in Fig.\ref{fig7}(a). Here, $a_0=\sqrt{\frac{\mu}{t_0+t_J}}$ is the length of minor axes of the elliptical Fermi surfaces.
Therefore, for $\theta=0$, as $t_J$ increases from zero, the critical current for $\alpha=0$ decays more rapidly than that for $\alpha=0.25\pi$, as shown by the black curves in Figs. \ref{fig6}(a) and \ref{fig6}(c).
%Therefore, the critical current decays more rapidly at $\alpha=0$ than in the case of $\alpha=0.25\pi$. This feature of the critical current for $\theta=0$ at small $t_J$ is reflected by the black lines in Fig. \ref{fig5}(a).
The rapid decay and oscillation of the critical current for $\theta=0$ can also be found in the ferromagnetic Josephson junctions (see Fig.D1(c) in Appendix D). In contrast to the ferromagnetic case, the oscillations here are intense and irregular. This results from the anisotropy of AM. Overall, a larger $t_J$ leads to more pronounced anisotropy and more prominent oscillation, which is consistent with the variation of the critical current with the orientation angle in Figs.\ref{fig5}(b)-(d).

For $\theta=0.5\pi$, the Andreev reflection at the two interfaces converts the incoming spin-down (up) electron from the AM into the spin-down (up) hole entering the AM, and vice versa. The Andreev reflection processes are schematically shown in Figs.\ref{fig7}(c) and (d) by the circles on the gray dashed lines labeled ``1(2)''. The electrons and the Andreev reflected holes have the same spin. They occupy the identical ellipses with mutually parallel major axes, as shown in Figs.\ref{fig7}(c) and (d). The wave vectors of the electrons and the holes are equal.  In this case, the critical current decays slowly, especially for small $t_J$. This feature of the critical current is reflected by the blue line in Fig.\ref{fig6}(a). For $\theta=0.5\pi$, the slow decay of the critical current mainly arises from the mismatch between the Fermi wave vectors in STSs and AM caused by the increase of $t_J$.

In addition, the critical current for $\theta=0$ exhibits anomalous behavior when $t_J$ approaches $1$. This can be understood by examining the changes in the Fermi surfaces and the group velocities of particles. For $\alpha=0$, energies of the electrons and holes can be written as $E_{e\uparrow(\downarrow)}=(t_0+(-)t_J)k_x^2+(t_0-(+)t_J)k_y^2-\mu$ and $E_{h\uparrow(\downarrow)}=-[(t_0+(-)t_J)k_x^2+(t_0-(+)t_J)k_y^2]+\mu$, respectively. The $x$ components of the group velocities are given by
\begin{eqnarray}
\begin{split}
v_{e\uparrow(\downarrow)x}&=2[t_0+(-) t_J]k_x,\\
v_{h\uparrow(\downarrow)x}&=-2[t_0+(-) t_J]k_x.\\
\end{split}
\end{eqnarray}
When $t_{J}\rightarrow 1(t_0)$, the $x$ components of the velocities of the spin-down electrons and the spin-down holes become zero. For $\theta=0$, the Andreev reflection involving the spin-down electrons and the spin-up holes or the spin-up electrons and the spin-down holes ceases to occur, and the critical current drops to zero as shown by the black line in Fig. \ref{fig6}(a). However, for $\theta=0.5\pi$, the Andreev reflection involving the spin-up electrons and the spin-up holes can still occur. The critical current takes a finite value in the limit $t_J\rightarrow 1$, as shown by the blue line in Fig.\ref{fig6}(a).

It is worth pointing out that the mechanism by which the critical current becomes zero in the altermagnetic junctions is different from that in the ferromagnetic junctions. When the magnetization in the ferromagnet increases to become equal to the chemical potential, the ferromagnet becomes a half-metal. The Fermi surfaces for spin-up electrons and spin-up holes vanish as shown by the Hamiltonian in Appendix D. The Andreev reflection for $\theta=0$, involving the electrons and holes with opposite spins, cannot occur. The critical current decreases to zero. However, the four crossing points of the Fermi surfaces in the AM remain unchanged as $t_J$ increases, as shown in Fig.\ref{fig7}. The Fermi surfaces for the spin-up and spin-down particles always exist. The vanishing of the critical current for $\theta=0$ and $\alpha=0$ is a result of the group velocities of spin-down electrons and spin-down holes becoming zero.

For $\alpha=0.25\pi$, the energies of the electrons and holes can be written as $E_{e\uparrow(\downarrow)}=t_0(k_x^2+k_y^2)+(-)2t_Jk_xk_y-\mu$ and $E_{h\uparrow(\downarrow)}=-[t_0(k_x^2+k_y^2)+(-)2t_Jk_xk_y]+\mu$, respectively. The $x$ components of the group velocities are given by
\begin{eqnarray}
\begin{split}
v_{e\uparrow(\downarrow)x}&=2[t_0k_x+(-) t_J k_y],\\
v_{h\uparrow(\downarrow)x}&=-2[t_0k_x+(-) t_J k_y].
\end{split}
\end{eqnarray}
When $t_{J}\rightarrow 1$, the Fermi surfaces for the particles are described by $\vert k_x\pm k_y\vert=\sqrt{\frac{\mu}{t_0}}$. The $x$ components of the group velocities gradually increase to $2\sqrt{t_0\mu}$. The increase of the group velocities can offset the decay caused by the unequal wave vectors, and the critical current for $\theta=0$ begins to rise as shown in Fig.\ref{fig6}(c).

From the above discussions and the numerical results, we find that the critical current strongly depends on the form of the altermagnetism in the AM. It shows distinct behaviors for the $d_{x^2-y^2}$-wave AM and the $d_{xy}$-wave AM although they can be related by a rotation of $\frac{\pi}{4}$ angle. From the perspective of symmetry, the two STSs do not have the rotational symmetry of $\frac{\pi}{4}$ angle. Even if the rotation can transform the $d_{x^2-y^2}$-wave AM into the $d_{xy}$-wave AM, we cannot obtain the relation $I(\alpha=0)=I(\alpha=\frac{\pi}{4})$. During the transformation, other parameters of the junctions also change.

In the preceding discussion of the numerical results, we conducted a comparison between ferromagnetic junctions and altermagnetic junctions. To the best of our knowledge, antiferromagnetic Josephson junctions involving STSs remain unexplored. As a type of collinear magnet, an antiferromagnet possesses a spin-degenerate band structure. For example, the two-dimensional $G$-type antiferromagnet discussed in Ref.[\onlinecite{Baltz}] exhibits this property. Furthermore, its energy bands are isotropic in momentum space.
These characteristics make antiferromagnets similar to conventional non-magnetic metals. Consequently, antiferromagnets are fundamentally different from AMs, it is difficult to detect the direction of
the ${\bf d}$-vector from an STS/antiferromagnet/STS junction.
%AMs possess a spin-splitting band structure and their energy bands are anisotropic in momentum space. Therefore, when STSs are involved, it is believed that altermagnetic junctions also exhibit different Josephson coupling compared to antiferromagnetic junctions.

\section{\label{sec5}Conclusions}

We propose a Josephson scheme to detect the direction of the $\bf{d}$-vector in STS. Due to the involvement of AM with the zero-net macroscopic magnetization, our scheme can eliminate the possible influence of the stray field on the $\bf{d}$-vector and help determine the intrinsic direction of the vector. We demonstrate the validity of our scheme by calculating the Josephson effects using the chiral superconducting state as an example. We find that the CPRs of the STS/AM/STS junctions strongly depend on the direction of the $\bf{d}$-vector and the orientation angle of AM. The $0$-$\pi$ transition can be realized by rotating the $\bf{d}$-vector or changing the crystallographic orientation of AM. The dependences of the critical current on the direction of the $\bf{d}$-vector, the orientation angle and the altermagnetism strength are also clarified. Our results possess the significance in the identification of the spin-triplet superconductivity and the design of the quantum devices.

Regarding experimental feasibility, first, the control of crystallographic orientation is a mature experimental technique \cite{56}; second, the typical altermagnetic strength in altermagnets is $t_J\sim0.5t_0$ \cite{56}. Our numerical results show that the direction of the intrinsic $\bf{d}$-vector, particularly the difference between the in-plane and out-of-plane $\bf{d}$-vectors, is well reflected when the parameters $\alpha=0.125\pi$ and $t_J=0.5$ (in units of $t_0$) are adopted. As discussed above, these parameter values are experimentally achievable. Therefore, our proposed Josephson junctions and the associated numerical results are experimentally feasible.

\section*{\label{sec6}ACKNOWLEDGMENTS}
This work was financially supported
by the National Natural Science Foundation of China
under Grants Nos. 12474046 and 12374034,
the National Key R and D Program of China (Grant No. 2024YFA1409002),
the Quantum Science and Technology-National Science and
Technology Major Project (2021ZD0302403),
and the projects ZR2023MA005 and ZR2022QA110 supported by Shandong Provincial Natural Science Foundation. We acknowledge
the High-performance Computing Platform of Peking University
for providing computational resources.

\appendix

%\section*{\label{sec5} APPENDIX}

\section{The Discretized Hamiltonian} \label{A}

\setcounter{equation}{0}
\setcounter{figure}{0}
\renewcommand{\theequation}{A\arabic{equation}}
\renewcommand{\thefigure}{A\arabic{figure}}

For the STS and AM segments, we elaborate on the explicit forms of
$\check{H}_{0}$
$\check{H}_{x}$,
$\check{H}_{y}$,
$\check{H}_{xy}$,
and $\check{H}_{x\bar{y}}$
in conjunction with their corresponding Eqs.(\ref{6}) and Eqs.(\ref{7}).

For the left (right) STS,
\begin{eqnarray}
	\check{H}_{0}^{L(R)} &= &\begin{pmatrix}
		\epsilon_0 & 0 & 0 & 0 \\[5pt]
		0 & \epsilon_0 & 0 & 0 \\[5pt]
		0 & 0 & -\epsilon_0 & 0 \\[5pt]
		0 & 0 & 0 & -\epsilon_0
	\end{pmatrix} \label{figA1},\\
%\end{equation}
%\begin{equation}
%	\setlength{\arraycolsep}{4pt}
	\check{H}_{x}^{L(R)} & = & \begin{pmatrix}
		-\frac{t_0}{a^2}  & 0 & \frac{\Delta_{\uparrow\uparrow}^{L(R)}}{2i a} & \frac{\Delta_{\uparrow\downarrow}^{L(R)}}{2i a} \\[6pt]
		0 & -\frac{t_0}{a^2} & \frac{\Delta_{\downarrow\uparrow}^{L(R)}}{2i a} & \frac{\Delta_{\downarrow\downarrow}^{L(R)}}{2i a}\\[6pt]
		\frac{\Delta_{\uparrow\uparrow}^{L(R)*}}{2i a} & \frac{\Delta_{\downarrow\uparrow}^{L(R)*}}{2i a} & \frac{t_0}{a^2}  & 0 \\[6pt]
		\frac{\Delta_{\uparrow\downarrow}^{L(R)*}}{2i a} & \frac{\Delta_{\downarrow\downarrow}^{L(R)*}}{2i a} & 0 & \frac{t_0}{a^2}
	\end{pmatrix}
	\label{figA2},
\end{eqnarray}
and
\begin{equation}
	\check{H}_{y}^{L(R)} = \begin{pmatrix}
		-\frac{t_0}{a^2}  & 0 & \frac{\Delta_{\uparrow\uparrow}^{L(R)}}{2a} & \frac{\Delta_{\uparrow\downarrow}^{L(R)}}{2a} \\[6pt]
		0 & -\frac{t_0}{a^2} & \frac{\Delta_{\downarrow\uparrow}^{L(R)}}{2a} & \frac{\Delta_{\downarrow\downarrow}^{L(R)}}{2a}\\[6pt]
		\frac{\Delta_{\uparrow\uparrow}^{L(R)*}}{-2a} & \frac{\Delta_{\downarrow\uparrow}^{L(R)*}}{-2a} & \frac{t_0}{a^2}  & 0 \\[6pt]
		\frac{\Delta_{\uparrow\downarrow}^{L(R)*}}{-2a} & \frac{\Delta_{\downarrow\downarrow}^{L(R)*}}{-2a} & 0 & \frac{t_0}{a^2}
	\end{pmatrix}
	\label{figA3},
\end{equation}
where $\epsilon_0=\frac{4t_0}{a^2} - \mu $, $\Delta_{\uparrow\uparrow}^{L(R)}=\Delta(-n_{L(R)x}+in_{L(R)y})e^{i\varphi_{l(r)}}$,
$\Delta_{\uparrow\downarrow}^{L(R)}=\Delta_{\downarrow\uparrow}^{L(R)}= \Delta n_{L(R)z} e^{i\varphi_{l(r)}}$ and $\Delta_{\downarrow\downarrow}^{L(R)}=\Delta(n_{L(R)x}+in_{L(R)y})e^{i\varphi_{l(r)}}$.

For the AM,
 \begin{eqnarray}
 	\check{H}_{0}^{AM} & = & \begin{pmatrix}
 		\epsilon_0 & 0 & 0 & 0 \\[5pt]
 		0 & \epsilon_0 & 0 & 0 \\[5pt]
 		0 & 0 & -\epsilon_0 & 0 \\[5pt]
 		0 & 0 & 0 & -\epsilon_0
 	\end{pmatrix}	\label{figA4},\\
% \end{equation}
% \begin{equation}
% 		\setlength{\arraycolsep}{0.5pt}
 %	\begin{split}
 		{H}_{x}^{AM} &=&
 		\left(\begin{array}{cccccc}
           -t_{0J+}&0&0 &0 \\[7pt]
 			0&-t_{0J-}&0&0\\[7pt]
 			0&0&t_{0J+}&0 \\[7pt]
 		0	& 0&0 & t_{0J-}
 		\end{array}\right)
 %	\end{split}	
 \label{figA5},\\
% \end{equation}
%\begin{equation}
		\setlength{\arraycolsep}{0.5pt}
%	\begin{split}
		{H}_{y}^{AM}&=
		&\left(\begin{array}{cccccc}
		-t_{0J-}  & 0 & 0 & 0\\[7pt]
		0 & -t_{0J+} & 0 & 0\\[7pt]
		0& 0& t_{0J-}  & 0 \\[7pt]
		0 & 0 & 0 & t_{0J+}
		\end{array}\right)
%	\end{split}
	\label{figA6},
\end{eqnarray}
\begin{equation}
	\setlength{\arraycolsep}{0.5pt}
	\check{H}_{xy}^{AM} = \begin{pmatrix}
		\frac{-t_J\sin(2\alpha)}{2a^2}  & 0 & 0 & 0\\[7pt]
		0 & \frac{t_J\sin(2\alpha)}{2a^2} & 0 & 0\\[7pt]
		0& 0& \frac{t_J\sin(2\alpha)}{2a^2}  & 0 \\[7pt]
		0 & 0 & 0 & \frac{-t_J\sin(2\alpha)}{2a^2}
	\end{pmatrix}
	\label{figA7},
\end{equation}
and
\begin{equation}
	\setlength{\arraycolsep}{0.5pt}
	\check{H}_{x\bar{y}}^{AM} = \begin{pmatrix}
		\frac{t_J\sin(2\alpha)}{2a^2}  & 0 & 0 & 0\\[7pt]
		0 & \frac{-t_J\sin(2\alpha)}{2a^2} & 0 & 0\\[7pt]
		0& 0& \frac{-t_J\sin(2\alpha)}{2a^2}  & 0 \\[7pt]
		0 & 0 & 0 & \frac{t_J\sin(2\alpha)}{2a^2}
	\end{pmatrix}
	\label{figA8},
\end{equation}
with $t_{0J\pm} = \frac{t_0 \pm t_J\cos(2\alpha)}{a^2}$.

\section{The derivation of the Green's functions} \label{B}
\setcounter{equation}{0}
\setcounter{figure}{0}
\renewcommand{\theequation}{B\arabic{equation}}
\renewcommand{\thefigure}{B\arabic{figure}}

After discretizing the model into lattice sites, we define
$H^{L(R)}_{11}$
as the Hamiltonian of an isolated slice for the left (right) STS, and
$H^{L(R)}_{12}$
as the hopping Hamiltonian from one slice to its adjacent slice on the right. Accordingly, we are able to derive the M\"obius transformation matrix
\begin{equation}
	\setlength{\arraycolsep}{6pt}
	X_L = \begin{pmatrix}
		0&(H^{L}_{12})^{-1}\\[5pt]
	-(H^{L}_{12})^{\dagger} &[(E+i\gamma)-H^{L}_{11}](H^{L}_{12})^{-1}
	\end{pmatrix}
	\label{figB1},
\end{equation}
with $\gamma$ being a small positive quantity. Employing the transformation matrix
$U_L$ to diagonalize the matrix $X_L$, we obtain
\begin{equation}
	U_{L}^{-1}X_{L}U_{L}=\text{diag}(\lambda_{L1},\lambda_{L2},\lambda_{L3},...)
		\label{figB2},
	\end{equation}
where $\lambda_{Ln}$ with $n=1,2,3...$ represent the eigenvalues of matrix $ X_{L}$,
ordered from left to right according to the increasing magnitude of their absolute values, i.e., $|\lambda_{L1}| < |\lambda_{L2}| < |\lambda_{L3}|<...$.
We consider the matrix $ U_L$ to be partitioned into the following $2 \times 2$  block form
\begin{equation}
		\setlength{\arraycolsep}{6pt}
	U_{L}=\begin{pmatrix}
			U_{L}^{11}&	U_{L}^{12}\\[5pt]
			U_{L}^{21}&U_{L}^{22}
	\end{pmatrix}
	\label{figB3}.
\end{equation}
Then, the surface Green's function for the left STS can be given by
$g^{r}_{L}=U_{L}^{12}(U_{L}^{22})^{-1}$. Similarly, by expressing the M\"obius transformation matrix $X_R$ and the matrix $U_R$
for the right STS segment, we can also obtain its surface Green's function as
$g^{r}_{R}=U_{R}^{12}(U_{R}^{22})^{-1}$.

We use the recursive formula to calculate the  retarded Green's function of AM, and the Green's function at its rightmost end is given by
\begin{equation}
G^{Rr}_{AM}(E,N_x)=[E-H^{11}_{AM}-\tilde{T}g^{r}_{R}(E)\tilde{T}^\dagger]^{-1}
	\label{figB4},
\end{equation}
where $H^{11}_{AM}$ denotes the Hamiltonian of the isolated slice of the center AM and $\tilde{T}=1_{N_y\times N_y}\otimes \check{T}$. The Green's function for the \textit{i}th slice can be derived via a recursive algorithm as:
\begin{equation}	G^{Rr}_{AM}(E,i)=[E-H^{11}_{AM}-H^{12}_{AM}G^{Rr}_{AM}(E,i+1)H^{21}_{AM}]^{-1}
	\label{figB5},
\end{equation}
where $H^{12}_{AM}$ denotes the hopping Hamiltonian to the adjacent slice on the right-hand side. The retarded Green's function for the leftmost slice can be derived as
\begin{equation}
		\begin{split}
	&G^{r}_{AM}(E,1)=\\
&[E-H^{11}_{AM}-\tilde{T}g^{r}_{R}(E)\tilde{T}^\dagger-H^{12}_{AM}G^{Rr}_{AM}(E,2)H^{21}_{AM}]^{-1}]^{-1}
	\label{figB6}.
		\end{split}
\end{equation}
The complete advanced Green's function is derived from the following relation
$G^{a}_{AM}(E)=[G^{r}_{AM}(E)]^\dagger$. Next, the full lesser Green's function for the leftmost slice of the AM can be expressed as
$G^{<}(E)=-f(E)[G^{r}_{AM}-G^{a}_{AM}]$. On this basis, the Josephson current can be derived through calculation.

\section{Proof of zero-net macroscopic magnetization} \label{C}
\setcounter{equation}{0}
\setcounter{figure}{0}
\renewcommand{\theequation}{C\arabic{equation}}
\renewcommand{\thefigure}{C\arabic{figure}}

According to Eq.(\ref{5}), the total Hamiltonian of AM can be written as
\begin{eqnarray}
h=h_{\uparrow}+h_{\downarrow},
\end{eqnarray}
with the spin-dependent Hamiltonians
\begin{eqnarray}
\begin{split}
h_{\uparrow}=&\sum_{{\bf{k}}}c_{\bf{k}\uparrow}^{+}[t_0k^2-\mu+t_{J}(k_{x}^2-k_{y}^2)\cos{2\alpha}\\
&+2k_x k_y\sin{2\alpha}]c_{\bf{k}\uparrow},\\
h_{\downarrow}=&\sum_{{\bf{k}}}c_{\bf{k}\downarrow}^{+}[t_0k^2-\mu-t_{J}(k_{x}^2-k_{y}^2)\cos{2\alpha}\\
&-2k_x k_y\sin{2\alpha}]c_{\bf{k}\downarrow}.
\end{split}
\end{eqnarray}
We introduce the operation $Y=T\mathcal{R}$. The unitary rotation $\mathcal{R}=D_{z}(\frac{\pi}{2})R_{z}(\frac{\pi}{2})$ allows the AM to undergo a spatial rotation of $\frac{\pi}{2}$ about the $z$-axis, accompanied by a spin rotation of $\frac{\pi}{2}$. The transformations of this operation on the annihilation operators are as follows,
\begin{eqnarray}
\begin{split}
Y c_{(k_x,ky)\uparrow}Y^{-1}&=c_{(k_y,-k_x)\downarrow}e^{i\frac{\pi}{2}},\\
Y c_{(k_x,ky)\downarrow}Y^{-1}&=-c_{(k_y,-k_x)\uparrow}e^{i\frac{\pi}{2}}
\end{split}
\end{eqnarray}
Then, the transformation of the spin-dependent Hamiltonians can be expressed as
\begin{eqnarray}
Y h_{\uparrow}Y^{-1}=h_{\downarrow},~~~Y h_{\downarrow}Y^{-1}=h_{\uparrow}.
\end{eqnarray}
As the sum of the spin-dependent Hamiltonians, the total Hamiltonian is invariant under the transformation, i.e., $Y hY^{-1}=h$. Therefore, if the net macroscopic magnetization is defined as $M=M_{\uparrow}-M_{\downarrow}$ in the AM, it is also invariant under the transformation. Here, $M_{\uparrow}$ is the magnetization from the spin-up electrons while $M_{\downarrow}$ is the magnetization from the spin-down electrons. On the other hand, the time-reversal operator $T$ flips the magnetization. We will obtain the relation $M=-M$, which indicates that the net macroscopic magnetization must be zero.

\section{Josephson effects for STS/ferromagnet/STS junctions} \label{D}
\setcounter{equation}{0}
\setcounter{figure}{0}
\renewcommand{\theequation}{D\arabic{equation}}
\renewcommand{\thefigure}{D\arabic{figure}}

The Hamiltonian of a ferromagnet can be written as
\begin{eqnarray}
H_{F}=\sum_k\psi_{F\bf{k}}^{\dagger}\check{H}_{F}({\bf{k}})\psi_{F\bf{k}},\label{HF}
\end{eqnarray}
with $\psi_{F\bf{k}}=(c_{F\bf{k}\uparrow},c_{F\bf{k}\downarrow},c_{F\bf{-k}\uparrow}^{\dagger},c_{F\bf{-k}\downarrow}^{\dagger})^T$ and the BdG Hamiltonian
\begin{eqnarray}
	\check{H}_{F}(\bf{k})=\left(\begin{array}{cc}
		h_{F}(\bf{k})&0\\
		0&-h_{F}^*(-\bf{k})
	\end{array}\right).\label{HFM}
\end{eqnarray}
Here $ h_{F}(\textbf{k}) = [t_{0}(k_{x}^{2} + k_{y}^{2}) - \mu_F]\sigma_0 + t_{F}\sigma_{z}$. The parameter $t_{F}$ characterizes the magnitude of the magnetization and $\mu_{F}$ is the chemical potential. The direction of the magnetization in the ferromagnet is fixed along the $+z$-axis. The spins of particles in the ferromagnet are parallel to the $z$-axis, which is the same as those in the AM.

\begin{figure*}[!htb]
	\centerline{\includegraphics[width=2\columnwidth]{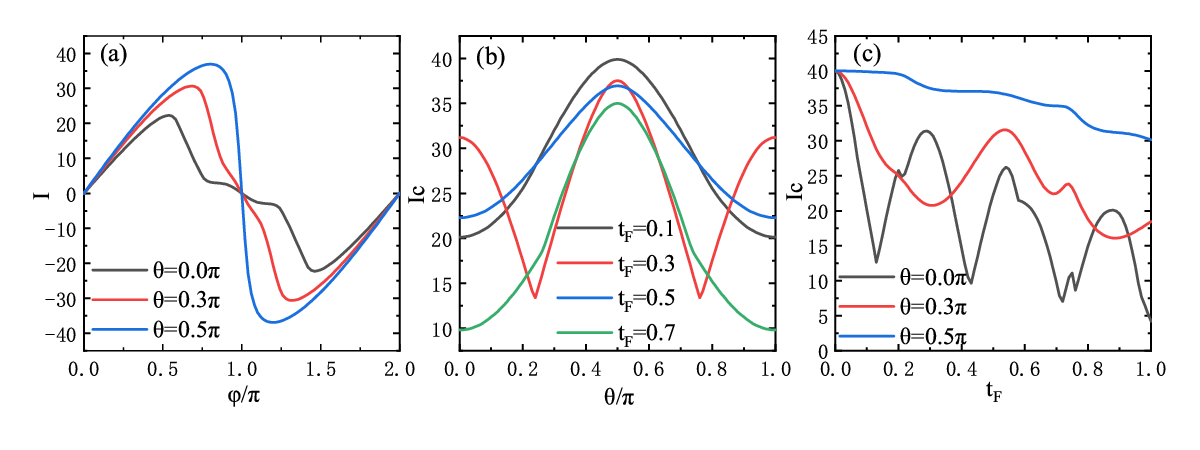}}
	\caption{(a) The CPRs for different polar angles of the ${\bf{d}}$-vectors. (b) The critical current as functions of the polar angles for different values of $t_F$. (c) The critical current as functions of the magnitude of magnetization for different polar angles. In (a), the magnitude of magnetization is take as $t_F=0.5$.}\label{figD1}
\end{figure*}
Using the Green's function method, the Josephson current can be calculated. For comparison with the results of the altermagnetic Josephson junctions, we adopt identical parameters, i.e., $\mu_F=2$, $t_F=0.5$, $N_x=N_y=10$, $a=1$, $t=1$ and $\Delta_0=0.01$. The numerical results for the ferromagnetic Josephson junctions are independent of the orientation of the ferromagnet since its Fermi surfaces are isotropic.

Fig.\ref{figD1}(a) shows the CPRs for different polar angles of the ${\bf{d}}$-vectors. For $\theta=0.5\pi$, the CPR is of the $\sin{\phi}$ form. When the polar angle changes from $0.5\pi$ to $0$, the $\sin{2\phi}$-type current gradually increases. Fig.\ref{figD1}(b) shows the critical current as a function of the polar angle. For the specific values of $t_F$ (e.g., $t_F=0.3$), the $0$-$\pi$ transition occurs as shown by the dips on the red line. Fig.\ref{figD1}(c) shows the variations of the critical current with $t_F$. The rapid decay and oscillations of the critical current for $\theta\ne0.5\pi$ are caused by the unequal wave vectors of electrons and the Andreev reflected holes. Especially for $\theta=0$, as $t_F$ continues to increase to $t_F=\mu_F$, the ferromagnet becomes a half-metal. The Fermi surfaces for the spin-up electrons and the spin-up holes vanish. The Andreev reflection, which correlates electrons and holes with opposite spins, does not occur. The critical current in this case vanishes.


\section*{REFERENCES}

\begin{thebibliography}{}
	\bibitem{1}
A. P. Mackenzie and Y. Maeno, The superconductivity of Sr$_2$RuO$_4$ and the physics of spin-triplet pairing,
\href{https://doi.org/10.1103/RevModPhys.75.657}{Rev. Mod. Phys. \textbf{75}, 657 (2003)}.

\bibitem{2}
Y.-S. Li, M. Garst, J. Schmalian, S. Ghosh, N. Kikugawa, D. A. Sokolov, C. W. Hicks, F. Jerzembeck, M. S. Ikeda, Z. Hu, B. J. Ramshaw, A. W. Rost, M. Nicklas, and A. P. Mackenzie, Elastocaloric determination of the phase diagram of Sr$_2$RuO$_4$,
\href{https://doi.org/10.1038/s41586-022-04820-z}{Nature \textbf{607}, 276 (2022)}.


\bibitem{3}
R. Khasanov, A. Ramires, V. Grinenko, I. Shipulin, N. Kikugawa, D. A. Sokolov, J. A. Krieger, T. J. Hicken, Y. Maeno, H. Luetkens, and Z. Guguchia, In-plane magnetic penetration depth in Sr$_2$RuO$_4$: Muon-spin rotation and relaxation study, \href{https://doi.org/10.1103/PhysRevLett.131.236001}{Phys. Rev. Lett. \textbf{131}, 236001 (2023)}.


\bibitem{4}
J. F. Landaeta, K. Semeniuk, J. Aretz, K. R. Shirer, D. A. Sokolov, N. Kikugawa, Y. Maeno, I. Bonalde, J. Schmalian, A. P. Mackenzie, and E. Hassinger, Evidence for vertical line nodes in Sr$_2$RuO$_4$ from nonlocal electrodynamics,
\href{https://doi.org/10.1103/PhysRevB.110.L100503}{Phys. Rev. B.\textbf{110}, L100503 (2024).}



\bibitem{5}
V. Grinenko, S. Ghosh, R. Sarkar, J.-C. Orain, A. Nikitin, M. Elender, D. Das, Z. Guguchia, F. Br$\ddot{\text{u}}$ckner, M. E. Barber, J. Park, N. Kikugawa, D. A. Sokolov, J. S. Bobowski, T. Miyoshi, Y. Maeno, A. P. Mackenzie, H. Luetkens, C. W. Hicks, and H.-H. Klauss, Split superconducting and time-reversal symmetry-breaking transitions in Sr$_2$RuO$_4$ under stress, \href{https://doi.org/10.1038/s41567-021-01182-7}{Nat. Phys. \textbf{17}, 748 (2021)}.

\bibitem{6}
S. Ghosh, A. Shekhter, F. Jerzembeck, N. Kikugawa, S. A. Sokolov, M. Brando, A. P. Mackenzie, C. W. Hicks, and B. J. Ramshaw, Thermodynamic evidence for a two-component superconducting order parameter in Sr$_2$RuO$_4$, \href{https://doi.org/10.1038/s41567-020-1032-4}{Nat. Phys. \textbf{17}, 199 (2021)}.

\bibitem{7}
S. Benhabib, C. Lupien, I. Paul, L. Berges, M. Dion, M. Nardone, A. Zitouni, Z. Q. Mao, Y. Maeno, A. Georges, L. Tailefer, and C. Proust, Ultrasound evidence for a two-component superconducting order parameter in Sr$_2$RuO$_4$,
\href{https://doi.org/10.1038/s41567-020-1033-3}{Nat. Phys. \textbf{17}, 194 (2021)}.

\bibitem{8}
R. Balian and N. R. Werthamer, Superconductivity with pairs in a relative p wave,
\href{https://doi.org/10.1103/PhysRev.131.1553}{Phys. Rev. \textbf{131}, 1553 (1963)}.

\bibitem{9}
M. Sato, Topological properties of spin-triplet superconductors and Fermi surface topology in the normal state,
\href{https://doi.org/10.1103/PhysRevB.79.214526}{Phys. Rev. B \textbf{79}, 214526 (2009)}.

\bibitem{10}
M. Sato, Nodal structure of superconductors with time-reversal invariance and Z$_2$ topological number, \href{https://doi.org/10.1103/PhysRevB.73.214502}{Phys. Rev. B \textbf{73}, 214502 (2006)}.

\bibitem{11}
Y. Maeno, A. Ikeda, and G. Mattoni, Thirty years of puzzling superconductivity in Sr$_2$RuO$_4$, \href{https://doi.org/10.1038/s41567-024-02656-0}{Nat. Phys. \textbf{20}, 1712 (2024)}.

\bibitem{12}
G. M. Luke, Y. Fudamoto, K. M. Kojima, M. I. Larkin, J. Merrin, B. Nachumi, Y. J. Uemura, Y. Maeno, Z. Q. Mao, Y. Mori, H. Nakamura, and M. Sigrist, Time-reversal symmetry-breaking superconductivity in Sr$_2$RuO$_4$,
\href{https://doi.org/10.1038/29038}{Nature \textbf{394}, 558 (1998)}.

\bibitem{13}
H. Matsumura, Y. Takahashi, K. Kinjo, S. Kitagawa, K. Ishida, Y. Tokunaga, H. Sakai, S. Kambe, A. Nakamura, et al., \textit{b}-axis and \textit{c}-axis Knight shift measurements in the superconducting state on ultraclean $\text{UTe}_2$ with \(T_c=2.1\ \text{K}\),
\href{https://doi.org/10.1103/PhysRevB.111.174526}{Phys. Rev. B \textbf{111}, 174526 (2025)}.

\bibitem{14}
K. Ishida, H. Mukuda, Y. Kitaoka, K. Asayama, Z. Q. Mao, Y. Mori, and Y. Maeno, Spin-triplet superconductivity in Sr$_2$RuO$_4$ identified by $^{17}$O Knight shift, \href{https://doi.org/10.1038/25315}{Nature \textbf{396}, 658 (1998)}.

\bibitem{15}
M. S. Anwar, R. Ishiguro, T. Nakamura, M. Yakabe, S. Yonezawa, H. Takayanagi, and Y. Maeno, Multicomponent order parameter superconductivity of Sr$_2$RuO$_4$ revealed by topological junctions,
\href{https://doi.org/10.1103/PhysRevB.95.224509}{Phys. Rev. B \textbf{95}, 224509 (2017)}.

\bibitem{16}
K. Ishida, H. Mukuda, Y. Kitaoka, Z. Q. Mao, Y. Mori, and Y. Maeno, Anisotropic superconducting gap in the spin-triplet superconductor Sr$_2$RuO$_4$: Evidence from a Ru-NQR study, \href{https://doi.org/10.1103/PhysRevLett.84.5387}{Phys. Rev. Lett. \textbf{84}, 5387 (2000)}.

\bibitem{17}
E. Hassinger, P. Bourgeois-Hope, H. Taniguchi, S. Ren$\acute{\text{e}}$ de Cotret, G. Grissonnanche, M. S. Anwar, Y. Maeno, N. Doiron-Leyraud, and L. Taillefer, Vertical line nodes in the superconducting gap structure of Sr$_2$RuO$_4$, \href{https://doi.org/10.1103/PhysRevX.7.011032}{Phys. Rev. X \textbf{7}, 011032 (2017)}.

\bibitem{18}
A. Pustogow, Y. Luo, A. Chronister, Y.-S. Su, D. A. Sokolov, F. Jerzembeck, A. P. Mackenzie, C. W. Hicks, N. Kikugawa, S. Raghu, E. D. Bauer, and S. E. Brown, Constraints on the superconducting order parameter in Sr$_2$RuO$_4$ from oxygen-17 nuclear magnetic resonance,
\href{https://doi.org/10.1038/s41586-019-1596-2}{Nature \textbf{574}, 72 (2019)}.

\bibitem{19}
M. Uchida, I. Sakuraba, M. Kawamura, M. Ide, K. S. Takahashi, Y. Tokura, and M. Kawasaki, Characterization of Sr$_2$RuO$_4$ Josephson junctions made of epitaxial films, \href{https://doi.org/10.1103/PhysRevB.101.035107}{Phys. Rev. B \textbf{101}, 035107 (2020)}.

\bibitem{20}
J. A. Duffy, S. M. Hayden, Y. Maeno, Z. Mao, J. Kulda, and G. J. McIntyre, Polarized-neutron scattering study of the Cooper-pair moment in Sr$_2$RuO$_4$, \href{https://doi.org/10.1103/PhysRevLett.85.5412}{Phys. Rev. Lett. \textbf{85}, 5412 (2000)}.

\bibitem{21}
K. D. Nelson, Z. Q. Mao, Y. Maeno, and Y. Liu, Odd-parity superconductivity in Sr$_2$RuO$_4$, \href{https://doi.org/10.1126/science.1103881}{Science \textbf{306}, 1151 (2004)}.

\bibitem{22}
A. N. Petsch, M. Zhu, M. Enderle, Z. Q. Mao, Y. Maeno, I. I. Mazin, and S. M. Hayden, Reduction of the spin susceptibility in the superconducting state of Sr$_2$RuO$_4$ observed by polarized neutron scattering, \href{https://doi.org/10.1103/PhysRevLett.125.217004}{Phys. Rev. Lett. \textbf{125}, 217004 (2020)}.

\bibitem{23}
S. Kashiwaya, K. Saitoh, H. Kashiwaya, M. Koyanagi, M. Sato, K. Yada, Y. Tanaka, and Y. Maeno, Time-reversal invariant superconductivity of Sr$_2$RuO$_4$ revealed by Josephson effects, \href{https://doi.org/10.1103/PhysRevB.100.094530}{Phys. Rev. B \textbf{100}, 094530 (2019)}.

\bibitem{24}
J. F. Dodaro, Z. Wang, and C. Kallin, Effects of deep superconducting gap minima and disorder on residual thermal transport in Sr$_2$RuO$_4$, \href{https://doi.org/10.1103/PhysRevB.98.214520}{Phys. Rev. B \textbf{98}, 214520 (2018)}.

\bibitem{25}
J.L. Zhang, W. Huang, M. Sigrist, and D.X. Lin, Interband interference effects at the edge of a multiband chiral $p$-wave superconductor, \href{https://doi.org/10.1103/PhysRevB.96.224504}{Phys. Rev. B \textbf{96}, 224504 (2017)}.

\bibitem{26}
L.D. Zhang, W. Huang, F. Yang, and H. Yao, Superconducting pairing in Sr$_2$RuO$_4$ from weak to intermediate coupling, \href{https://doi.org/10.1103/PhysRevB.97.060510}{Phys. Rev. B \textbf{97}, 060510(R) (2018)}.

\bibitem{27}
S. B. Etter, A. Bouhon, and M. Sigrist, Spontaneous surface flux pattern in chiral $p$-wave superconductors,  \href{https://doi.org/10.1103/PhysRevB.97.064510}{Phys. Rev. B \textbf{97}, 064510 (2018)}.

\bibitem{28}
W. Huang and H. Yao, Possible three-dimensional nematic odd-parity superconductivity in Sr$_2$RuO$_4$, \href{https://doi.org/10.1103/PhysRevLett.121.157002}{Phys. Rev. Lett. \textbf{121}, 157002 (2018)}.

\bibitem{29}
L. A. B. Olde Olthof, S.I. Suzuki, A. A. Golubov, M. Kunieda, S. Yonezawa, Y. Maeno, and Y. Tanaka, Theory of tunneling spectroscopy of normal metal/ferromagnet/spin-triplet superconductor junctions, \href{https://doi.org/10.1103/PhysRevB.98.014508}{Phys. Rev. B \textbf{98}, 014508 (2018)}.

\bibitem{30}
Y. Imai, K. Wakabayashi, and M. Sigrist, Thermal Hall conductivity in the spin-triplet superconductor with broken time-reversal symmetry, \href{https://doi.org/10.1103/PhysRevB.95.024516}{Phys. Rev. B \textbf{95}, 024516 (2017)}.

\bibitem{31}
K. Kawai, K. Yada, Y. Tanaka, Y. Asano, A. A. Golubov, and S. Kashiwaya, Josephson effect in a multiorbital model for Sr$_2$RuO$_4$, \href{https://doi.org/10.1103/PhysRevB.95.174518}{Phys. Rev. B \textbf{95}, 174518 (2017)}.

\bibitem{32}
T. Fukumoto, K. Taguchi, S. Kobayashi, and Y. Tanaka, Theory of tunneling conductance of anomalous Rashba metal/superconductor junctions,
\href{https://doi.org/10.1103/PhysRevB.92.144514}{Phys. Rev. B \textbf{92}, 144514 (2015)}.

\bibitem{33}
S.I. Suzuki, M. Sato, and Y. Tanaka, Identifying possible pairing states in Sr$_2$RuO$_4$ by tunneling spectroscopy,  \href{https://doi.org/10.1103/PhysRevB.101.054505}{Phys. Rev. B \textbf{101}, 054505 (2020)}.

\bibitem{34}
S.I. Suzuki, S. Ikegaya, and A. A. Golubov, Destruction of surface states of ($d_{zx}+id_{yz}$)-wave superconductor by surface roughness: Application to Sr$_2$RuO$_4$, \href{https://doi.org/10.1103/PhysRevResearch.4.L042020}{Phys. Rev. Res. \textbf{4}, L042020 (2022)}.

\bibitem{35}
S. Ikegaya, S.I. Suzuki, Y. Tanaka, and D. Manske, Proposal for identifying possible even-parity superconducting states in Sr$_2$RuO$_4$ using planar tunneling spectroscopy, \href{https://doi.org/10.1103/PhysRevResearch.3.L032062}{Phys. Rev. Res. \textbf{3}, L032062 (2021)}.

\bibitem{36}
O. Maistrenko, C. Autieri, G. Livanas, P. Gentile, A. Romano, C. Noce, D. Manske, and M. Cuoco, Inverse proximity effects at spin-triplet superconductor-ferromagnet interface, \href{https://doi.org/10.1103/PhysRevResearch.3.033008}{Phys. Rev. Res. \textbf{3}, 033008 (2021)}.

\bibitem{37}
Q. Cheng, Q. Yan, and Q.-F. Sun, Spin-triplet superconductor$-$uantum anomalous Hall insulator$-$spin-triplet superconductor Josephson junctions: 0-$\pi$ transition, $\varphi_0$ phase, and switching effects,
 \href{https://doi.org/10.1103/PhysRevB.104.134514}{Phys. Rev. B \textbf{104}, 134514 (2021)}.

\bibitem{Annett}
J. F. Annett, B. L. Gy$\ddot{\text{o}}$rffy, G. Litak, and K. I. wysoki$\acute{\text{n}}$ski, Magnetic field induced rotation of the $\bf{d}$-vector in the spin-triplet superconductor Sr$_2$RuO$_4$
\href{https://doi.org/10.1103/PhysRevB.78.054511}{Phys. Rev. B \textbf{78}, 054511 (2008)}.

\bibitem{Murakawa}
H. Murakawa, K. Ishida, K. Kitagawa, Z. Q. Mao, and Y. Maeno, Measurement of the $^{101}$Ru-Knight Shift of Superconducting Sr$_2$RuO$_4$ in a Parallel Magnetic Field,
\href{https://doi.org/10.1103/PhysRevLett.93.167004}{Phys.Rev.Lett
\textbf{93},167004(2004)}.

\bibitem{KIshida}
K. Ishida, M. Manago, K. Kinjo, and Y. Maeno, Reduction of the $^17$O Knight Shift in the Superconducting Stateand the Heat-up Effect by NMR Pulses on Sr$_2$RuO$_4$, J. Phys. Soc. Jpn. \textbf{89}, 034712 (2020).

 \bibitem{38}
 I. Mazin, and The PRX Editors, Editorial: Altermagnetism$-$New Punch Line of Fundamental Magnetism,
 \href{https://doi.org/10.1103/PhysRevX.12.040002}{Phys. Rev. X \textbf{12}, 040002 (2022)}.

 \bibitem{39}
 L. $\check{\text{S}}$mejkal, J. Sinova, and T. Jungwirth, Beyond Conventional Ferromagnetism and Antiferromagnetism: A Phase with Nonrelativistic Spin and Crystal Rotation Symmetry,
 \href{https://doi.org/10.1103/PhysRevX.12.031042}{Phys. Rev. X \textbf{12}, 031042 (2022)}.

 \bibitem{41}
 L. $\check{\text{S}}$mejkal, J. Sinova, and T. Jungwirth, Emerging Research Landscape of Altermagnetism,
 \href{https://doi.org/10.1103/PhysRevX.12.040501}{Phys. Rev. X \textbf{12}, 040501 (2022)}.

 \bibitem{42}
 J. Krempask\'{y}, L. \v{S}mejkal, S. W. D'Souza \textit{et al.}, Altermagnetic lifting of Kramers spin degeneracy,
 \href{https://doi.org/10.1038/s41586-023-06907-7}{Nature \textbf{626}, 517 (2024)}.


 \bibitem{43}
 X. Zhou, W. Feng, R.W. Zhang, L. \v{S}mejkal, J. Sinova, Y. Mokrousov, and Y. Yao, Crystal Thermal Transport in Altermagnetic $RuO_{2}$,
 \href{https://doi.org/10.1103/PhysRevLett.132.056701}{Phys. Rev. Lett. \textbf{132}, 056701 (2024)}.

 \bibitem{49}
Y.Y. Li, and S.B. Zhang, Floating edge bands in the Bernevig-Hughes-Zhang model with altermagnetism,
\href{https://doi.org/10.1103/PhysRevB.111.045106}{Phys. Rev. B \textbf{111}, 045106 (2025)}.


\bibitem{50}
Z.M. Wang, Y. Zhang, S.B. Zhang, J.H. Sun, E. Dagotto, D.H. Xu, and L.H. Hu, Spin-Orbital Altermagnetism,
\href{https://doi.org/10.1103/cjzw-j4v7}{Phys. Rev. Lett.
\textbf{135}, 176705 (2025)}.

\bibitem{51}
H.P. Sun, S.B. Zhang, C.A. Li, and B. Trauzettel, Tunable second harmonic in altermagnetic Josephson junctions,
\href{https://doi.org/10.1103/PhysRevB.111.165406}{Phys. Rev. B \textbf{111}, 165406 (2025)}.

\bibitem{52}
H.J. Lin, S.B. Zhang, H.Z. Lu, and X. C. Xie, Coulomb Drag in Altermagnets,
\href{https://doi.org/10.1103/PhysRevLett.134.136301}{Phys. Rev. Lett. \textbf{134}, 136301 (2025)}.

\bibitem{53}
X. Zhu, X. Huo, S. Feng, S.B. Zhang, S. A. Yang, and H. Guo, Design of Altermagnetic Models from Spin Clusters,
\href{https://doi.org/10.1103/PhysRevLett.134.166701}{Phys. Rev. Lett. \textbf{134}, 166701 (2025)}.

\bibitem{Ghorashi}
S. A. A. Ghorashi, T. L. Hughes, and J. Cano, Altermagnetic Routes to Majorana Modes in Zero Net Magnetization,
\href{https://doi.org/10.1103/PhysRevLett.133.106601}{Phys. Rev. Lett.
\textbf{133}, 106601 (2024)}.

\bibitem{YXLi}
Y.X. Li and C.C. Liu, Majorana corner modes and tunable patterns in an altermagnet heterostructure,
\href{https://doi.org/10.1103/PhysRevB.108.205410}{Phys. Rev. B \textbf{108}, 205410 (2023)}.

\bibitem{Giil}
H.G. Giil and J. Linder, Superconductor-altermagnet memory functionality without stray fields,
 \href{https://doi.org/10.1103/PhysRevB.109.134511}{Phys. Rev. B \textbf{109}, 134511 (2024)}.

\bibitem{add1}
W. Chen, X. Zhou, W.-K. Lou, and K. Chang, Magneto-optical conductivity and circular dichroism in d-wave altermagnets,
\href{https://journals.aps.org/prb/abstract/10.1103/PhysRevB.111.064428}{Phys. Rev. B \textbf{111}, 064428 (2025)}.

\bibitem{add2}
Y.-F. Sun, Y. Mao, Y.-C. Zhuang, and Q.-F. Sun,
Tunneling magnetoresistance effect in altermagnets,
\href{https://journals.aps.org/prb/abstract/10.1103/t8b5-l859}{Phys. Rev. B \textbf{112}, 094411 (2025)}.

\bibitem{add3}
Y.-H. Wan, P.-Y. Liu, and Q.-F. Sun,
Interplay of altermagnetic order and Wilson mass in the Dirac equation:
Helical edge states without time-reversal symmetry,
\href{https://journals.aps.org/prb/abstract/10.1103/s6pj-495v}{Phys. Rev. B \textbf{112}, 115412 (2025)}.


\bibitem{Herasymchuk}
A. Herasymchuk, K. B. Hallberg, E. W. Hodt, J. Linder, E. V. Gorbar, and P. Sukhachov, Electric and spin current vortices in altermagnets, \href{https://doi.org/10.1103/3sw6-y8vf }{Phys. Rev. B \textbf{112}, L220404 (2025)}.

\bibitem{Yarmohammadi}
M. Yarmohammadi, U. Z$\ddot{\text{u}}$licke, J. Berakdar, J. Linder, and J. K. Freericks, Anisotropic light-tailored RKKY interaction in two-dimensional $d$-wave altermagnets, \href{https://doi.org/10.1103/k3xb-8pts}{Phys. Rev. B \textbf{111}, 224412 (2025)}.

\bibitem{Tjernshaugen}
J. B. Tjernshaugen, M. Amundsen, and J. Linder, Crossed Andreev reflection revealed by self-consistent Keldysh-Usadel formalism, \href{https://doi.org/10.1103/PhysRevB.110.224502}{Phys. Rev. B \textbf{110}, 224502 (2024)}.

\bibitem{Sukhachov}
P. O. Sukhachov, E. W. Hodt, and J. Linder, Thermoelectric effect in altermagnet-superconductor junctions,  \href{https://doi.org/10.1103/PhysRevB.110.094508}{Phys. Rev. B \textbf{110}, 094508 (2024)}.


\bibitem{add4}
P.-Y. Liu, Y.-H. Wan, and Q.-F. Sun, Emergence of net chirality in a two-dimensional Dirac fermion system with altermagnetic mass,  \href{https://journals.aps.org/prb/abstract/10.1103/k6th-nyt9}{Phys. Rev. B \textbf{113}, L041402 (2026)}.



\bibitem{Hodt}
E.W. Hodt and J. Linder, Spin pumping in an altermagnet/normal-metal bilayer,
 \href{https://doi.org/10.1103/PhysRevB.109.174438}{Phys. Rev. B \textbf{109}, 174438 (2024)}.

\bibitem{Amundsen}
M. Amundsen, A. Brataas, and J. Linder, RKKY interaction in Rashba altermagnets,
\href{https://doi.org/10.1103/PhysRevB.110.054427}{Phys. Rev. B \textbf{110}, 054427 (2024)}.


\bibitem{Papaj}
M. Papaj, Andreev reflection at the altermagnet-superconductor interface,   \href{https://doi.org/10.1103/PhysRevB.108.L060508}{Phys. Rev. B \textbf{108}, L060508 (2023)}.

 \bibitem{46}
 C. Sun, A. Brataas, and J. Linder, Andreev reflection in altermagnets,
 \href{https://doi.org/10.1103/PhysRevB.108.054511}{Phys. Rev. B \textbf{108}, 054511 (2023)}.


 \bibitem{47}
 C. W. J. Beenakker, and T. Vakhtel, Phase-shifted Andreev levels in an altermagnet Josephson junction,
 \href{https://doi.org/10.1103/PhysRevB.108.075425}{Phys. Rev. B \textbf{108}, 075425 (2023)}.

 \bibitem{57}
J. A. Ouassou, A. Brataas, and J. Linder, dc Josephson Effect in Altermagnets,
\href{https://doi.org/10.1103/PhysRevLett.131.076003}{Phys. Rev. Lett. \textbf{131}, 076003 (2023)}.


\bibitem{45}
 Q. Cheng, and Q.-F. Sun, Orientation-dependent Josephson effect in spin-singlet superconductor/altermagnet/spin-triplet superconductor junctions,
 \href{https://doi.org/10.1103/PhysRevB.109.024517}{Phys. Rev. B \textbf{109}, 024517 (2024)}.

\bibitem{56}
Q. Cheng, Y. Mao, and Q.-F. Sun, Field-free Josephson diode effect in altermagnet/normal metal/altermagnet junctions,
\href{https://doi.org/10.1103/PhysRevB.110.014518}{Phys. Rev. B \textbf{110}, 014518 (2024)}.

\bibitem{Banerjee}
S. Banerjee and M. S. Scheurer, Altermagnetic superconducting diode effect, \href{https://doi.org/10.1103/PhysRevB.110.024503}{Phys. Rev. B \textbf{110}, 024503 (2024)}.

\bibitem{55}
Y. Fukaya, K. Maeda, K. Yada, J. Cayao, Y. Tanaka, and B. Lu, Josephson effect and odd-frequency pairing in superconducting junctions with unconventional magnets,
\href{https://doi.org/10.1103/PhysRevB.111.064502}{Phys. Rev. B \textbf{111}, 064502 (2025)}.


\bibitem{Maeda}
K. Maeda, Y. Fukaya, K. Yada, B. Lu, Y. Tanaka, and J. Cayao, Classification of pair symmetries in superconductors with unconventional magnetism, \href{https://doi.org/10.1103/PhysRevB.111.144508}{Phys. Rev. B \textbf{111}, 144508 (2025)}.

\bibitem{Sunjpcm}
Q.-F. Sun and X.C. Xie, Quantum transport through a graphene nanoribbon Csuperconductor junction,
\href{https://doi.org/10.1088/0953-8984/21/34/344204}{J. Phys.: Condens. Matter \textbf{21}, 344204 (2009)}.


\bibitem{YHLi}
Y.H. Li, J. Song, J. Liu, H. Jiang, Q.-F. Sun, and X.C. Xie, Doubled Shapiro steps in a topological Josephson junction, \href{https://doi.org/10.1103/PhysRevB.97.045423}{Phys. Rev. B \textbf{97}, 045423 (2018)}.


\bibitem{Kastening}
B. Kastening, D. Morr, D. Manske, and K. Bennemann, Novel Josephson Effect in Triplet-Superconductor-Ferromagnet-Triplet-Superconductor Junctions, \href{https://doi.org/10.1103/PhysRevLett.96.047009}{Phys. Rev. Lett. \textbf{96}, 047009 (2006)}.

\bibitem{Brydon1}
P.M.R. Brydon, B. Kastening, D.K. Morr, and D. Manske, Interplay of ferromagnetism and triplet superconductivity in a Josephson junction,   \href{https://doi.org/10.1103/PhysRevB.77.104504}{Phys. Rev. B \textbf{77}, 104504 (2008)}.

\bibitem{Brydon2}
P.M.R. Brydon and D. Manske, $0$-$\pi$ Transition in Magnetic Triplet Superconductor Josephson Junctions, \href{https://doi.org/10.1103/PhysRevLett.103.147001}{Phys. Rev. Lett. \textbf{103}, 147001 (2009)}.


\bibitem{Brydon4}
P.M.R. Brydon, Y. Asano, and C. Timm, Spin Josephson effect with a single superconductor, \href{https://doi.org/10.1103/PhysRevB.83.180504}{Phys. Rev. B \textbf{83}, 180504(R) (2011)}.


\bibitem{Jungwirth}
T. Jungwirth, X. Marti, P. Wadley, and J. Wunderlich, Antiferromagnetic spintronics,
\href{https://www.nature.com/articles/nnano.2016.18}{Nat. Nanotechnol. \textbf{11}, 231 (2016)}.

\bibitem{Baltz}
V. Baltz, A. Manchon, M. Tsoi, T. Moriyama, T. Ono, and Y. Teerkovnyak, Antiferromagnetic spintronics, \href{https://journals.aps.org/rmp/pdf/10.1103/RevModPhys.90.015005} {Rev. Mod. Phys. \textbf{90}, 015005 (2018)}.
\end{thebibliography}
\end{document}